\documentclass[aps,pre,preprint,showpacs,eqsecnum]{revtex4}
\usepackage{graphicx,epsf,amssymb,amsmath}

\begin{document}

\title{Consistent description of kinetics and hydrodynamics of weakly nonequilibrium processes in simple liquids}

\author{B. Markiv, I. Omelyan, M. Tokarchuk}

\affiliation{Institute for Condensed Matter Physics 
of the National Academy of Sciences of Ukraine,
1 Svientsitskii Srt., 79011 Lviv, Ukraine}

\begin{abstract}
The generalized transport equations for a consistent description of kinetic
and hydrodynamic processes in dense gases and liquids are considered.
The inner structure of the generalized transport kernels for these equations is established. It is shown how in this approach to obtain the transport equation of molecular hydrodynamics. For the model potential of interaction presented as a sum of the hard spheres potential and certain long-range potential a spectrum of collective modes in the system is investigated.
\end{abstract}



\maketitle

\section{Introduction}\label{sec:1}

A number of
investigations~\cite{Temper,118,Get,98,Bogol,74,For,March,Res,77,zub5,zub6,klim2,100,101,101a,101b,102,tok2,zub4,tok,zub2,luzzi1,luzzi2,luzzi3,luzzi4,tsyt,mar}
was devoted to the problem of constructing a consistent
description of kinetic and hydrodynamic processes in dense gases,
liquids, and plasma.

In dense gases and liquids (simple and molecular) as well as in dense plasma there is no small parameter and
the characteristic time of interparticle correlations
is comparable with those for the one-particle distribution
function. This means that during the particles collision process the many-particles
correlations related to local mass, momentum and energy
conservation laws, underlying the hydrodynamic description of a
system, can not be neglected. In this connection the local
conservation laws impose some restrictions on the kinetic processes.
Their role is especially important at high densities, when
the interaction between a separate group of particles and other ones can
not be neglected. This indicates a close connection between the kinetic
and hydrodynamic processes in dense gases, liquids and plasma.
Such peculiarities of transport processes should be displayed in the behaviour
of collective excitations spectrum, time correlation functions (in particular, the dynamic structure factor)
as well as the generalized transport coefficient in the region of intermediate values of wave vector $\vec{k}$ and frequency $\omega$.
Another important issue remains for investigation when the nonequilibrium processes are considered
during a long time (small $\omega$) at a small spatial scale (large $\vec{k}$) between particles and when the features of interparticle interaction character are manifested.
In particular, it is important to note the results of comparison of experimental data on neutron scattering  and molecular dynamics ones~\cite{van}. For all values of $\vec{k}$ in the region of small $\omega$,
deviations are observed unlike for the region of high frequencies (short time of observation).
Whereas at small $\vec{k}$ the behaviour can be explained by collective effects, obviously, the dynamics of particles scattering with a momentum exchange should be taken into account at large $\vec{k}$. This is very important from the point of view of experimental investigations on neutron scattering~\cite{Temper,74,Bul}.

For instance, the
importance of taking into account the kinetic processes connected
with irreversible collision processes at the scale of short-ranged
interparticle interactions was pointed out in~\cite{deSchep}. The short-wavelength collective modes in liquids were
investi\-gated therein on the basis of the linearized kinetic
equation of the revised Enskog theory for the hard
spheres model.

In this paper we investigate a spectrum of collective excitations
within a consistent description of kinetic and hydrodynamic
processes in a system when potential of interaction between
particles consists of two parts: the hard spheres potential and a
long-range part. In the second section we present basic transport equations of a consistent description of kinetics and hydrodynamics of weakly nonequilibrium processes obtained by means of the Zubarev nonequilibrium statistical operator (NSO) method \cite{zub2}.
The memory functions entering these equations  are calculated in section~3.
In the fourth section based on the transport equations of a consistent description of kinetics and hydrodynamics we obtain the equations of molecular hydrodynamics. In section~5 the latter are used for investigation of spectrum of collective excitations in the system with potential of interaction modelled by a sum of the potential of hard spheres and a certain long-range potential.

\section{Kinetic equations for weakly nonequilibrium states}\label{sec:2}

Using the ideas of papers~\cite{zub5,zub6} the nonequilibrium
statistical operator for a consistent description of kinetic and
hydrodynamic processes for a system of classical interacting
particles was obtained in~\cite{zub4,tok} by means of
the NSO method.
For this purpose the following approximation for quasiequilibrium distribution function
being the functional of the reduced-description parameters $\langle\hat{n}_{\vec{k}}(\vec{p})\rangle^t$ and $\langle\hat{h}_{\vec{k}}^{int}\rangle^t$, was used:
\begin{eqnarray}
\label{math/2.3}
\varrho_q(x^N;t)=\varrho_0(x^N)\Bigl\{1+{\sum_{\vec{k}}}'
\langle\hat{h}_{\vec{k}}^{int}\rangle^t\Phi_{hh}^{-1}(\vec{k})
\hat{h}^{int}_{\vec{k}}+{\sum_{\vec{k}}}'\int d\vec{p}\int
d\vec{p}'
\langle\hat{n}_{\vec{k}}(\vec{p})\rangle^t\Phi_{\vec{k}}^{-1}(\vec{p},\vec{p}')
\hat{n}_{\vec{k}}(\vec{p}')\Bigr\}.\nonumber\\
\end{eqnarray}
Here, $\varrho_0(x^N)$ is an equilibrium distribution function, ${\sum_{\vec{k}}}'=\sum_{\vec{k}(\vec{k}\neq 0)}$, with $\vec{k}$ being a wave vector.
\begin{eqnarray}
\label{math/2.4} \hat{n}_{\vec{k}}(\vec{p})=\int d\vec{r}e^{-i\vec{k}\cdot\vec{r}}\hat{n}_1(\vec{r},\vec{p})
\end{eqnarray}
are the Fourier-components of microscopic phase density of
particles number,
\begin{eqnarray}
\label{math/2.5}
\hat{h}_{\vec{k}}^{int}=\hat{\varepsilon}_{\vec{k}}^{int}
-\langle\hat{\varepsilon}_{\vec{k}}^{int}\hat{n}_{-\vec{k}}\rangle_0S_2^{-1}(k)
\hat{n}_{\vec{k}}
\end{eqnarray}
are the Fourier-components of the potential part of the enthalpy density,
$\hat{\varepsilon}_{\vec{k}}^{int}=\frac{1}{2}\sum_{l\neq
j=1}^N\Phi(|\vec{r}_{lj}|)e^{-i\vec{k}\cdot\vec{r}_j}$ and
$\hat{n}_{\vec{k}}=\sum_{l=1}^Ne^{-i\vec{k}\cdot\vec{r}_l}$
are the Fourier-components of the potential energy and
particles number densities, respectively. $\Phi_{hh}^{-1}(\vec{k})$ is the
function inverse to the equilibrium correlation function
$\Phi_{hh}(\vec{k})=\langle\hat{h}_{\vec{k}}^{int}
\hat{h}_{-\vec{k}}^{int}\rangle_0,$
$\langle\ldots\rangle_0=\int d\Gamma_N\ldots\varrho_0(x^N)$,
$\langle\hat{n}_{\vec{k}}(\vec{p})\rangle_0=0(\vec{k}\neq 0)$.
$\Phi_{\vec{k}}^{-1}(\vec{p},\vec{p}')$ is the function inverse to
\[\Phi_{\vec{k}}(\vec{p},\vec{p}')=\langle\hat{n}_{\vec{k}}(\vec{p})
\hat{n}_{-\vec{k}}(\vec{p}')\rangle_0=n\delta(\vec{p}-\vec{p}')f_0(p')
+n^2f_0(p)f_0(p')h_2(\vec{k}).
\]
It  equals to
\begin{eqnarray}
\label{math/2.7}
\Phi_{\vec{k}}^{-1}(\vec{p},\vec{p}')=\frac{\delta(\vec{p}-\vec{p}')}{nf_0(p')}-c_2(k),
\end{eqnarray}
where $n=N/V$, $f_0(p)=\left({\beta}/{2\pi
m}\right)^{3/2}\exp(-\beta{p^2}/{2m})$ is the Maxwellian
distribution, $\beta=1/k_BT$ with $T$ being an equilibrium value of temperature. $c_2(k)$ is the direct correlation function
related to correlation function $h_2(k)$:
$h_2(k)=c_2(k)[1-nc_2(k)]^{-1}$.
$S_2(k)=\langle\hat{n}_{\vec{k}}\hat{n}_{-\vec{k}}\rangle_0$ is
the static structure factor. It is important to note that dynamic
variables $\hat{h}_{\vec{k}}^{int}$ and
$\hat{n}_{\vec{k}}(\vec{p})$ in the distribution (\ref{math/2.3})
are orthogonal in the sense that
$\langle\hat{h}_{\vec{k}}^{int}\hat{n}_{\vec{k}}(\vec{p})\rangle_0=0$.

In the approximation (\ref{math/2.3}), within the framework of the Zubarev NSO method \cite{zub1,zub2}, the nonequilibrium distribution function $\varrho(x^N;t)$ has the following form~\cite{zub4,tok}:
\begin{eqnarray}
\label{math/2.8}
\lefteqn{\varrho(x^N;t)=\varrho_0(x^N)\Bigl\{1+{\sum_{\vec{k}}}'\int
d\vec{p}\int d\vec{p}'\langle\hat{n}_{\vec{k}}(\vec{p})\rangle^t
\Phi_{\vec{k}}^{-1}(\vec{p},\vec{p}')\hat{n}_{-\vec{k}}(\vec{p}')}
\\&&\mbox{}+{\sum_{\vec{k}}}'\langle\hat{h}^{int}_{\vec{k}}\rangle^t
\Phi_{hh}^{-1}(\vec{k})\hat{h}^{int}_{-\vec{k}}-{\sum_{\vec{k}}}'\int
d\vec{p}\int
d\vec{p}'\int_{-\infty}^te^{\varepsilon(t'-t)}T_0(t,t')I_n(-\vec{k};\vec{p})
\Phi_{\vec{k}}^{-1}(\vec{p},\vec{p}')
\langle\hat{n}_{\vec{k}}(\vec{p}')\rangle^{t'}\nonumber
\\&&\mbox{}-{\sum_{\vec{k}}}'\int_{-\infty}^te^{\varepsilon(t'-t)}T_0(t,t')I_h^{int}(-\vec{k})
\Phi_{hh}^{-1}(\vec{k})
\langle\hat{h}^{int}_{\vec{k}}\rangle^{t'}\Bigr\},\nonumber
\end{eqnarray}
where
\begin{eqnarray}
\label{math/2.9} I_n(\vec{k};\vec{p})=(1-P_0)iL_N\hat{n}_{\vec{k}}(\vec{p})
=(1-P_0)\dot{\hat{n}}_{\vec{k}}(\vec{p}),\quad
I_h^{int}(\vec{k})=(1-P_0)iL_N\hat{h}^{int}_{\vec{k}}
=(1-P_0)\dot{\hat{h}}^{int}_{\vec{k}}\nonumber\\
\end{eqnarray}
are the generalized flows in linear approximation ($iL_N$ is the Liouville operator of a simple liquid).
$T_0(t,t')=\exp[(t-t')(1-P_0)iL_N]$ is the evolution operator
with regard to projection operator $P_0$ being a linear
approximation of the Mori projection operator constructed
on the orthogonal dynamic variables $\hat{n}_{\vec{k}}(\vec{p})$,
$\hat{h}_{\vec{k}}^{int}$~\cite{tok}:
\begin{eqnarray}
\label{math/2.10} P_0\hat{A}_{\vec{k}}={\sum_{\vec{k}}}'
\langle\hat{A}_{\vec{k}}\hat{h}^{int}_{-\vec{k}}\rangle_0\Phi_{hh}^{-1}(\vec{k})
\hat{h}^{int}_{\vec{k}}+{\sum_{\vec{k}}}'\int d\vec{p}\int
d\vec{p}'
\langle\hat{A}_{\vec{k}}\hat{n}_{-\vec{k}}(\vec{p})\rangle_0
\Phi_{\vec{k}}^{-1}(\vec{p},\vec{p}') \hat{n}_{\vec{k}}(\vec{p}').
\end{eqnarray}
It possesses the following properties $P_0P_0=P_0$, $P_0(1-P_0)=0$,
$P_0\hat{n}_{\vec{k}}(\vec{p})=\hat{n}_{\vec{k}}(\vec{p})$,
$P_0\hat{h}^{int}_{\vec{k}}=\hat{h}^{int}_{\vec{k}}$.

In view of its own structure, the nonequilibrium distribution
function (\ref{math/2.8}) is a functional of the
reduced-description parameters
$\langle\hat{n}_{\vec{k}}(\vec{p})\rangle^t$,
$\langle\hat{h}^{int}_{\vec{k}}\rangle^t$, dynamic variables
$\hat{n}_{\vec{k}}(\vec{p})$, $\hat{h}^{int}_{\vec{k}}$ along with
their generalized flows (\ref{math/2.9}). Using $\varrho(x^N;t)$
(\ref{math/2.8}) for the parameters of reduced description
$f_{\vec{k}}(\vec{p};t)=\langle\hat{n}_{\vec{k}}(\vec{p})\rangle^t$,
$h^{int}_{\vec{k}}(t)=\langle\hat{h}^{int}_{\vec{k}}\rangle^t$ we
can obtain the following set of equations~\cite{tok}:
\begin{eqnarray}
\label{math/2.11} \lefteqn{\frac{\partial}{\partial
t}f_{\vec{k}}(\vec{p};t)+\frac{i\vec{k}\cdot\vec{p}}{m}f_{\vec{k}}(\vec{p};t) =-\frac{i\vec{k}\cdot\vec{p}}{m}nf_0(p)c_2(k)\int
d\vec{p}'f_{\vec{k}}(\vec{p}';t)+i\Omega_{nh}(\vec{k};\vec{p})h^{int}_{\vec{k}}(t)}
\\&&\mbox{}-\int
d\vec{p}'\int_{-\infty}^{t}e^{\varepsilon(t'-t)}
\varphi_{nn}(\vec{k};\vec{p},\vec{p}';t,t')f_{\vec{k}}(\vec{p}';t')dt'
-\int_{-\infty}^{t}e^{\varepsilon(t'-t)}
\varphi_{nh}(\vec{k};\vec{p};t,t')h^{int}_{\vec{k}}(t')dt',\nonumber
\end{eqnarray}
\begin{eqnarray}
\label{math/2.12} \lefteqn{\frac{\partial}{\partial
t}h_{\vec{k}}^{int}(t)=\int d\vec{p}'i\Omega_{hn}(\vec{k};\vec{p}')f_{\vec{k}}(\vec{p}';t)}
\\&&\mbox{}-\int
d\vec{p}'\int_{-\infty}^{t}e^{\varepsilon(t'-t)}
\varphi_{hn}(\vec{k};\vec{p}';t,t')f_{\vec{k}}(\vec{p}';t')dt'
-\int_{-\infty}^{t}e^{\varepsilon(t'-t)}
\varphi_{hh}(\vec{k};t,t')h^{int}_{\vec{k}}(t')dt',\nonumber
\end{eqnarray}
Here, $i\Omega_{nh}(\vec{k};\vec{p})$, $i\Omega_{hn}(\vec{k};\vec{p})$ are the normalized static
correlation functions
\begin{eqnarray}
\label{math/2.13} i\Omega_{nh}(\vec{k};\vec{p})=
\langle\dot{\hat{n}}_{\vec{k}}(\vec{p})\hat{h}^{int}_{-\vec{k}}\rangle_0
\Phi_{hh}^{-1}(\vec{k}),\qquad i\Omega_{hn}(\vec{k};\vec{p})= \int
d\vec{p}'\langle\dot{\hat{h}}_{\vec{k}}^{int}\hat{n}_{-\vec{k}}(\vec{p}')\rangle_0
\Phi_{\vec{k}}^{-1}(\vec{p}',\vec{p})
\end{eqnarray}
and
\begin{eqnarray}
\label{math/2.15} &\varphi_{nn}(\vec{k};\vec{p},\vec{p}';t,t')=
\int d\vec{p}''\langle
I_{n}(\vec{k};\vec{p})T_0(t,t')I_{n}(-\vec{k};\vec{p}'')\rangle_0
\Phi_{\vec{k}}^{-1}(\vec{p}'',\vec{p}'),&\nonumber\\
&\varphi_{hn}(\vec{k};\vec{p};t,t')= \int d\vec{p}'\langle
I_{h}^{int}(\vec{k})T_0(t,t')I_{n}(-\vec{k};\vec{p}')\rangle_0
\Phi_{\vec{k}}^{-1}(\vec{p}',\vec{p}),&\\
&\varphi_{hn}(\vec{k};\vec{p};t,t')= \langle
I_{n}(\vec{k};\vec{p})T_0(t,t')I_{h}^{int}(-\vec{k})\rangle_0
\Phi_{hh}^{-1}(\vec{k}),&\nonumber\\
&\varphi_{hh}(\vec{k};\vec{p};t,t')= \langle
I_{h}^{int}(\vec{k})T_0(t,t')I_{h}^{int}(-\vec{k})\rangle_0
\Phi_{hh}^{-1}(\vec{k})&\nonumber
\end{eqnarray}
are the generalized transport kernels (memory functions)
describing kinetic and hydrodynamic processes. The set of
equations (\ref{math/2.11}) and (\ref{math/2.12}) is closed with
respect to the parameters of reduced description
$f_{\vec{k}}(\vec{p};t)$, $h^{int}_{\vec{k}}(t)$. If in this set
of equations one formally puts $\hat{h}^{int}_{\vec{k}}=0$, then
we obtain the kinetic equation for $f_{\vec{k}}(\vec{p};t)$:
\begin{eqnarray}
\label{math/2.19} \lefteqn{\frac{\partial}{\partial
t}f_{\vec{k}}(\vec{p};t)+\frac{i\vec{k}\cdot\vec{p}}{m}f_{\vec{k}}(\vec{p};t)=-\frac{i\vec{k}\cdot\vec{p}}{m}nf_0(p)c_2(k)\int
d\vec{p}'f_{\vec{k}}(\vec{p}';t)}\qquad
\\&&\mbox{}-\int
d\vec{p}'\int_{-\infty}^{t}e^{\varepsilon(t'-t)}
\varphi'_{nn}(\vec{k};\vec{p},\vec{p}';t,t')f_{\vec{k}}(\vec{p}';t')dt'.
\nonumber
\end{eqnarray}
This is true when the contribution form the potential energy is
considerably smaller than the averaged kinetic energy (e.g. in the
case of gases or weakly coupled liquids). The equation
(\ref{math/2.19}) was obtained for the first time by means of the
Mori projection operators method in~\cite{107,108,109}. Therein the
basic parameter of the reduced description was a nonequilibrium
one-particle distribution function $f_{\vec{k}}(\vec{p};t)$ which
corresponds to the microscopic phase density
$\hat{n}_{\vec{k}}(\vec{p})$ (the Klimontovich function). In this
case the memory function
$\varphi'_{nn}(\vec{k};\vec{p},\vec{p}';t,t')$ has the following
structure:
\begin{eqnarray}
\label{math/2.20} \varphi'_{nn}(\vec{k};\vec{p},\vec{p}';t,t')=
\int d\vec{p}''\langle
I_n^0(\vec{k},\vec{p})T'_0(t,t')I_n^0(-\vec{k},\vec{p}'')\rangle_0
\Phi_{\vec{k}}^{-1}(\vec{p}'',\vec{p}'),
\end{eqnarray}
where
$I_n^0(\vec{k},\vec{p})=(1-P'_0)\dot{\hat{n}}_{\vec{k}}(\vec{p})$
is the generalized flow, $P'_0$ is the Mori operator, introduced
in~\cite{107,108,109}
\begin{eqnarray}
\label{math/2.22} P'_0{\hat{A}}_{\vec{k}'}={\sum_{\vec{k}}}'\int
d\vec{p}\int
d\vec{p}'\langle{\hat{A}}_{\vec{k}'}{\hat{n}}_{-\vec{k}'}(\vec{p}')\rangle_0
\Phi_{\vec{k}}^{-1}(\vec{p}',\vec{p})\hat{n}_{\vec{k}}(\vec{p}),
\end{eqnarray}
and $T'_0(t,t')$ is the corresponding evolution operator with regard
to projection. Based on the kinetic equation (\ref{math/2.19}) the
investigations of the dynamic structure factor, transverse and
longitudinal current time correlation functions, diffusion and
viscosity coefficients for dense gases and liquids where carried
out~\cite{77,79,107,108,109,110,111,112,113,114,115,116,117,118,119}.
In particular, in Mazenko's papers the linearized Boltzmann-Enskog
equation was obtained by means of expansion of the memory
functions $\varphi'_{nn}(\vec{k};\vec{p},\vec{p}';t,t')$ in
density. For the case of weak coupling, the equation of
Fokker-Planck type was obtained. However, the main drawback of
kinetic equation (\ref{math/2.19}) consists in its inconsistency
with the total energy conservation law, especially for dense gases
and liquids, when the contribution of the potential energy into
the thermodynamic functions and transport coefficients is
determinant. In~\cite{118} this drawback was studied in detail.
There the investigations of the dynamic structure factor at
intermediate values of wave-vector $\vec{k}$ and frequency
$\omega$ for simple liquids were carried out by using the Mori
projection operators method for the reduced-description parameters
$\hat{n}_{\vec{k}}(\vec{p})$, $\hat{\varepsilon}_{\vec{k}}$.
Contrary to the transport equations presented in \cite{118}, our
set of equations (\ref{math/2.11}) and (\ref{math/2.12}) is
constructed on the orthogonal dynamic variables
$\hat{n}_{\vec{k}}(\vec{p})$, $\hat{h}^{int}_{\vec{k}}$.
Therefore, ``kinetic'' and ``hydrodynamic'' contributions are
separated and correlation between them is described by the
generalized memory functions (\ref{math/2.15}). It is important to
reveal their inner structure.

\section{Memory functions of
a consistent description of kinetic and hydrodynamic processes}\label{sec:3}

In order to study the structure of the memory functions
(\ref{math/2.15}) let us look at the form of the
corresponding generalized flows (\ref{math/2.9}) on which the memory
functions are built. In particular, let us take into consideration
the fact that
\begin{eqnarray}
\label{math/3.1} \dot{\hat{n}}_{\vec{k}}(\vec{p})=iL_N\hat{n}_{\vec{k}}(\vec{p})= -\frac{i\vec{k}}{m}\cdot\hat{\vec{\jmath}}_{\vec{k}}(\vec{p})
+\frac{\partial}{\partial \vec{p}}\cdot\vec{F}_{\vec{k}}(\vec{p}),
\end{eqnarray}
where
\begin{eqnarray}
\label{math/3.2} \hat{\vec{\jmath}}_{\vec{k}}(\vec{p})
=\sum_{j=1}^N\vec{p}_{j}e^{-i\vec{k}\cdot\vec{r}_j}\delta(\vec{p}-\vec{p}_j)
\end{eqnarray}
is the momentum density in the $(\vec{k},\vec{p})$ space and
\begin{eqnarray}
\label{math/3.3} \vec{F}_{\vec{k}}(\vec{p})
=\frac{1}{2}\sum_{j\neq l}\frac{\partial}{\partial
\vec{r}}_j\Phi(|\vec{r}_j-\vec{r}_l|)\delta(\vec{p}-\vec{p}_j)e^{-i\vec{k}\cdot\vec{r}_{j}}.
\end{eqnarray}
The action of the Mori projection operator $P_0$ on
$\dot{\hat{n}}_{\vec{k}}(\vec{p})$ can be presented as
\begin{eqnarray}
\label{math/3.4} P_0\dot{\hat{n}}_{\vec{k}}(\vec{p})
=-\frac{\beta}{m}\vec{p}\cdot\vec{\Phi}_{Fh}(\vec{k})\hat{h}^{int}_{\vec{k}}f_0(p)
-\frac{i\vec{k}}{m}\cdot\vec{p}\hat{n}_{\vec{k}}(\vec{p}),
\end{eqnarray}
where
\begin{eqnarray}
\label{math/3.5}
\vec{\Phi}_{Fh}(\vec{k})=\langle\vec{F}_{\vec{k}}\hat{h}_{\vec{k}}^{int}\rangle_0
\Phi_{hh}^{-1}(\vec{k}).
\end{eqnarray}
Taking into account (\ref{math/2.7}) we write down the memory
function $\varphi_{nn}(\vec{k};\vec{p},\vec{p}';t,t')$ in the
following form
\begin{eqnarray}
\label{math/3.6}
\lefteqn{\varphi_{nn}(\vec{k};\vec{p},\vec{p}';t,t') =\int
d\vec{p}''\langle
I_n(\vec{k};\vec{p})T_0(t,t')I_n(-\vec{k;\vec{p}''})\rangle_0
\left\{\frac{\delta(\vec{p}''-\vec{p}')}{nf_0(p')}-c_2(k)\right\}}
\\&&\mbox{}= \langle
I_n(\vec{k};\vec{p})T_0(t,t')I_n(-\vec{k;\vec{p}'})\rangle_0\frac{1}{nf_0(p')}
-\int d\vec{p}''\langle
I_n(\vec{k};\vec{p})T_0(t,t')I_n(-\vec{k;\vec{p}''})\rangle_0c_2(k).\nonumber
\end{eqnarray}
The second term in the right-hand side of (\ref{math/3.6}) is equal to zero, because
\begin{eqnarray}
\label{math/3.7} \int
d\vec{p}I_{n}(\vec{k};\vec{p})=0.
\end{eqnarray}
The transport kernel $\varphi_{nn}(\vec{k};\vec{p},\vec{p}';t,t')$
enters into the kinetic equation (\ref{math/2.11}) as the term\\
$\int
\varphi_{nn}(\vec{k};\vec{p},\vec{p}';t,t')f_{\vec{k}}(\vec{p}';t)d\vec{p}'$.
Taking into account (\ref{math/3.6}) and (\ref{math/3.7}) we can
write the last one in the form
\begin{eqnarray}
\label{math/3.8} \lefteqn{\int
d\vec{p}'\varphi_{nn}(\vec{k};\vec{p},\vec{p}';t,t')f_{\vec{k}}(\vec{p}';t)
=\int
d\vec{p}'\Biggl\{\bar{\varphi}_{\jmath\jmath}(\vec{k};\vec{p},\vec{p}';t,t')}
\\&&\mbox{}
-\frac{\partial}{\partial
\vec{p}}\cdot\varphi_{FF}(\vec{k};\vec{p},\vec{p}';t,t')
\cdot\left(\frac{\beta}{mnf_0(p')}\vec{p}'-\frac{\partial}{\partial
\vec{p}'}\right)\Biggr\}f_{\vec{k}}(\vec{p}';t')-\bar{\varphi}_{n\jmath}^{(2)}(\vec{k};\vec{p};t,t')
\cdot\langle\hat{\vec{\jmath}}_{\vec{k}}\rangle^t,\nonumber
\end{eqnarray}
where the second term has the structure of the generalized
Fokker-Planck operator containing the generalized friction
coefficient $\varphi_{FF}(\vec{k};\vec{p},\vec{p}';t,t')$ in the
spatially-impulse space.
\begin{eqnarray}
\label{math/3.9}
\lefteqn{\bar{\varphi}_{\jmath\jmath}(\vec{k};\vec{p},\vec{p}';t,t')=
\frac{\vec{k}}{m}\cdot\varphi_{\jmath\jmath}(\vec{k};\vec{p},\vec{p}';t,t')\cdot\frac{\vec{k}}{m}}
\\&&\mbox{}-
\frac{i\vec{k}}{m}\cdot\varphi_{\jmath
F}(\vec{k};\vec{p},\vec{p}';t,t')
\cdot\frac{\partial}{\partial\vec{p}'}+
\frac{\partial}{\partial\vec{p}}\cdot\varphi_{F\jmath}(\vec{k};\vec{p},\vec{p}';t,t')
\cdot\frac{i\vec{k}}{m}+\bar{\varphi}_{nn}^{(1)}(\vec{k};\vec{p},\vec{p}';t,t'),\nonumber
\end{eqnarray}
\begin{eqnarray}
\label{math/3.10} {\varphi}_{FF}(\vec{k};\vec{p},\vec{p}';t,t')=
\langle\vec{F}_{\vec{k}}(\vec{p})T_0(t,t')\vec{F}_{-\vec{k}}(\vec{p}')\rangle_0,
\qquad {\varphi}_{\jmath\jmath}(\vec{k};\vec{p},\vec{p}';t,t')=
\langle\hat{\vec{\jmath}}_{\vec{k}}(\vec{p})T_0(t,t')
\hat{\vec{\jmath}}_{-\vec{k}}(\vec{p}')\rangle_0,\nonumber\\
\end{eqnarray}
\begin{eqnarray*}
\label{math/3.12} {\varphi}_{\jmath
F}(\vec{k};\vec{p},\vec{p}';t,t')=
\langle\hat{\vec{\jmath}}_{\vec{k}}(\vec{p})T_0(t,t')
{\vec{F}}_{-\vec{k}}(\vec{p}')\rangle_0,\qquad
{\varphi}_{F\jmath}(\vec{k};\vec{p},\vec{p}';t,t')=
\langle{\vec{F}}_{\vec{k}}(\vec{p})T_0(t,t')
\hat{\vec{\jmath}}_{-\vec{k}}(\vec{p}')\rangle_0,
\end{eqnarray*}
moreover,
\begin{eqnarray}
\label{math/3.14}\int d\vec{p}\int d\vec{p}'
\varphi_{\jmath\jmath}(\vec{k};\vec{p},\vec{p}';t,t')=
D(\vec{k};t,t'),
\end{eqnarray}
is the generalized coefficient of diffusion of particles and
$\varphi_{\jmath\jmath}(\vec{k};\vec{p},\vec{p}';t,t')$ the
generalized diffusion coefficient in momentum space.
\begin{eqnarray}
\label{math/3.15}\int d\vec{p}\int d\vec{p}'
\varphi_{FF}(\vec{k};\vec{p},\vec{p}';t,t')= \xi(\vec{k};t,t')
\end{eqnarray}
is the generalized friction coefficient and
$\varphi_{FF}(\vec{k};\vec{p},\vec{p}';t,t')$ the generalized
friction coefficient in momentum space.

The structure of functions $\bar{\varphi}_{nn}^{(1)}(\vec{k};\vec{p},\vec{p}';t,t')$ and
$\varphi_{n\jmath}^{(2)}(\vec{k};\vec{p};t,t')$ is presented in the Appendix. In particular, they are determined by the time correlation functions built on the set of dynamic variables $\hat{n}_{\vec{k}}(\vec{p})$, $\hat{h}_{\vec k}^{int}$, $\hat{\vec{\jmath}}_{\vec{k}}(\vec{p})$ and
$\vec{F}_{\vec{k}}(\vec{p})$.

In the kinetic equation (\ref{math/2.11}) the transport kernel
$\varphi_{nh}(\vec{k};\vec{p};t,t')$ describes dynamic
correlations between the kinetic and hydrodynamic processes.
Performing the action of operators $(1-P_0)$ and $iL_N$
as well as
taking into account (\ref{math/3.4}) and
\begin{eqnarray}\label{math/3.20}
P_0\dot{\hat{h}}^{int}_{\vec{k}}=\vec{\Phi'}_{hF}(\vec{k})
\cdot\hat{\vec{\jmath}}_{\vec{k}}, \quad
\vec{\Phi'}_{hF}(\vec{k})=\langle\hat{h}_{\vec{k}}^{int}
\vec{F}_{-\vec{k}}\rangle_0\frac{\beta}{mn},
\end{eqnarray}
the kernel $\varphi_{nh}(\vec{k};\vec{p};t,t')$ can be presented
as
\begin{eqnarray}
\label{math/3.21} \lefteqn{\varphi_{nh}(\vec{k};\vec{p};t,t')
=-\frac{i\vec{k}}{m}\cdot\bar{\varphi}_{\jmath\dot{h}}(\vec{k};\vec{p};t,t')
+\frac{\partial}{\partial \vec{p}}
\cdot\bar{\varphi}_{F\dot{h}}(\vec{k};\vec{p};t,t')
+\frac{\beta}{m}\vec{p}\cdot\vec{\Phi}_{Fh}(\vec{k})f_0(p)
\bar{\varphi}_{h\dot{h}}(\vec{k};t,t')}\nonumber
\\&&\mbox{}+\frac{i\vec{k}}{m}\cdot\vec{p}
\bar{\varphi}_{n\dot{h}}(\vec{k};\vec{p};t,t') +\frac{i\vec{k}}{m}\cdot{\varphi}_{\jmath\jmath}(\vec{k};\vec{p};t,t')
\cdot\vec{\Phi}'_{hF}(\vec{k})-\frac{\partial}{\partial
\vec{p}}\cdot
\varphi_{F\jmath}(\vec{k};\vec{p};t,t')\vec{\Phi}'_{hF}(\vec{F})\nonumber
\\&&\mbox{}-\frac{\beta}{m}\vec{p}\cdot\vec{\Phi}_{Fh}(\vec{k})f_0(p)
\varphi_{h\jmath}(\vec{k};t,t')\cdot\vec{\Phi}'_{hF}(\vec{k})-\frac{i\vec{k}}{m}\cdot\vec{p}\varphi_{n\jmath}(\vec{k};\vec{p};t,t')
\vec{\Phi}'_{hF}(\vec{k}),
\end{eqnarray}
The structure of the correlation functions entering this equation is presented in the Appendix as well.
The correlation functions $\varphi_{\jmath\jmath}(\vec{k};\vec{p};t,t')$,
$\varphi_{F\jmath}(\vec{k};\vec{p};t,t')$,
$\varphi_{h\jmath}(\vec{k};\vec{p};t,t')$, and
$\varphi_{n\jmath}(\vec{k};\vec{p};t,t')$ have the structure
similar to (\ref{math/3.17}).

From the structure of the transport
kernels (\ref{math/3.8}), (\ref{math/3.9}), (\ref{math/3.16}),
(\ref{math/3.18}) and (\ref{math/3.20}) in the kinetic equation
(\ref{math/2.11}) for a nonequilibrium one-particle distribution
function one can see that the contributions of the hydrodynamic
processes are described, besides $\hat{h}_{\vec{k}}^{int}$, by the
moments $\int
d\vec{p}f_{\vec{k}}(\vec{p};t)=n_{\vec{k}}(t)=\langle\hat{n}_{\vec{k}}\rangle^t$ and
$\int d\vec{p}\vec{p}f_{\vec{k}}(\vec{p};t)=\langle
\hat{\vec{\jmath}}_{\vec{k}}\rangle^t$.

As in the case of
equation (\ref{math/2.11}), let us find the inner structure of
transport kernels in equation (\ref{math/2.12}) for the
average value of the potential part of the enthalpy. In particular, taking
into account (\ref{math/3.20}), for $\varphi_{hh}(\vec{k};t,t')$
we obtain:
\begin{eqnarray}
\label{math/3.24} \lefteqn{{\varphi}_{hh}(\vec{k};t,t')=
\bar{\varphi}_{\dot{h}\dot{h}}^{(0)}(\vec{k};t,t')
-\vec{\Phi}'_{hF}(\vec{k})\cdot\bar{\varphi}_{\jmath\dot{h}}^{(0)}(\vec{k};t,t')}
\\&&\mbox{}
-\bar{\varphi}_{\dot{h}\jmath}(\vec{k};t,t')\cdot
\vec{\Phi}'_{hF}(\vec{k})+\vec{\Phi}'_{hF}(\vec{k})\cdot
{\varphi}_{\jmath\jmath}^{(0)}(\vec{k};t,t')\cdot\vec{\Phi}_{Fh}(\vec{k}).\nonumber
\end{eqnarray}
\begin{eqnarray}
\label{math/3.25}
&\bar{\varphi}_{\dot{h}\dot{h}}^{(0)}(\vec{k};t,t')
=\langle\dot{\hat{h}}_{\vec{k}}^{int}T_0(t,t')\dot{\hat{h}}_{-\vec{k}}^{int}\rangle_0
\Phi_{hh}^{-1}(\vec{k}),
&\bar{\varphi}_{\jmath\dot{h}}^{(0)}(\vec{k};t,t')
=\langle\hat{\vec{\jmath}}_{\vec{k}}T_0(t,t')
\dot{\hat{h}}_{\vec{k}}^{int}\rangle_0
\Phi_{hh}^{-1}(\vec{k}),\\
&\bar{\varphi}_{\dot{h}\jmath}(\vec{k};t,t')
=\Phi_{hh}^{-1}(\vec{k})\langle\dot{\hat{h}}_{\vec{k}}^{int}T_0(t,t')
\hat{\vec{\jmath}}_{\vec{k}}\rangle_0 ,
&{\varphi}^{(0)}_{\jmath\jmath}(\vec{k};t,t')
=\langle\hat{\vec{\jmath}}_{\vec{k}}T_0(t,t')
\hat{\vec{\jmath}}_{\vec{k}}\rangle_0=D(\vec{k};t,t')\nonumber\\
\end{eqnarray}
that is, ${\varphi}^{(0)}_{\jmath\jmath}(\vec{k};t,t')$ is the generalized diffusion coefficient (\ref{math/3.14}), whereas, $\bar{\varphi}_{\dot{h}\dot{h}}^{(0)}(\vec{k};t,t')$ determines
a potential part of the generalized heat-conductivity
coefficient. Taking
into account (\ref{math/3.4}), (\ref{math/3.21}) the transport
kernel $\varphi_{hn}(\vec{k};\vec{p};t,t')$ entering
(\ref{math/2.12}) as $\int
d\vec{p}'\varphi_{hn}(\vec{k};\vec{p}';t,t')f_{\vec{k}}(\vec{p}';t')$
can be presented in the following way
\begin{eqnarray}
\label{math/3.26} \lefteqn{\int
d\vec{p}'\varphi_{hn}(\vec{k};\vec{p}';t,t')f_{\vec{k}}(\vec{p}';t')
=\frac{i\vec{k}}{m}\cdot\int
d\vec{p}'W_{hn}(\vec{k};\vec{p}';t,t')
\frac{1}{nf_0(p')}f_{\vec{k}}(\vec{p}';t)}
\\&&\mbox{}-\int d\vec{p}'W_{hF}(\vec{k};\vec{p}';t,t')
\frac{1}{nf_0(p')}\cdot\left(\beta\frac{\vec{p}'}{mn}-\frac{\partial}{\partial
\vec{p}'}\right)f_{\vec{k}}(\vec{p}';t')+W_{hj}(\vec{k};t,t')
\frac{\beta}{mn}\cdot\langle\hat{\vec{\jmath}}_{\vec{k}}\rangle^{t'},\nonumber
\end{eqnarray}
where
\begin{eqnarray*}
\lefteqn{W_{hn}(\vec{k};\vec{p}';t,t')=\varphi_{\dot{h}\jmath}(\vec{k};\vec{p}';t,t')}
\\&&\mbox{}-\varphi_{\dot{h}n}(\vec{k};\vec{p}';t,t')\cdot\vec{p}'
-\vec{\Phi}'_{hF}(\vec{k})\varphi_{\jmath\jmath}(\vec{k};\vec{p}';t,t')
+\vec{\Phi}'_{hF}(\vec{k})\varphi_{\jmath
n}(\vec{k};\vec{p}';t,t')\cdot\vec{p}',
\end{eqnarray*}
\begin{eqnarray*}
W_{hF}(\vec{k};\vec{p}';t,t')=\varphi_{\dot{h}F}(\vec{k};\vec{p}';t,t')
+\vec{\Phi}'_{hF}(\vec{k})\varphi_{\jmath
F}(\vec{k};\vec{p}';t,t'),
\end{eqnarray*}
\begin{eqnarray*}
W_{h\jmath}(\vec{k};t,t')=\varphi_{\dot{h}h}(\vec{k};t,t')\vec{\Phi}'_{hF}(\vec{k})
-\vec{\Phi}'_{hF}(\vec{k})\varphi_{\jmath
h}(\vec{k};t,t')\vec{\Phi}'_{hF}(\vec{k})
\end{eqnarray*}
are the transport kernels, formed by the time correlation
functions of type of (\ref{math/3.22}) and
(\ref{math/3.25}). Taking into account the structure of the memory
functions (\ref{math/3.8}), (\ref{math/3.21}), (\ref{math/3.24})
and (\ref{math/3.26}), we present the set of equations
(\ref{math/2.11}), (\ref{math/2.12}) in the form
\begin{eqnarray}
\label{math/3.27} \lefteqn{\frac{\partial}{\partial
t}f_{\vec{k}}(\vec{p};t)+\frac{i\vec{k}}{m}\cdot\vec{p}
f_{\vec{k}}(\vec{p};t)=}
\\&&\mbox{}-\frac{i\vec{k}}{m}\cdot\vec{p}nf_0(p)c_2(\vec{k})
\int d\vec{p}'f_{\vec{k}}(\vec{p}';t) +i\Omega_{nh}(\vec{k};\vec{p})h_{\vec{k}}^{int}(t)\nonumber
\\&&\mbox{}-\int d\vec{p}'\int_{-\infty}^{t}e^{\varepsilon(t'-t)}
\Bigl\{\varphi_{\jmath\jmath}(\vec{k};\vec{p},\vec{p}';t,t')\nonumber
\\&&\mbox{}-\frac{\partial}{\partial
\vec{p}}\cdot\varphi_{FF}(\vec{k};\vec{p},\vec{p}';t,t')
\cdot\left(\frac{\beta\vec{p}'}{mnf_0(p')}-\frac{\partial}{\partial
\vec{p}'}\right)\Bigr\}f_{\vec{k}}(\vec{p}';t')dt'\nonumber
\\&&\mbox{}+\int_{-\infty}^{t}e^{\varepsilon(t'-t)}
\varphi^{(2)}_{n\jmath}(\vec{k};\vec{p},;t,t')
\cdot\vec{\jmath}_{\vec{k}}(t)\nonumber
\\&&\mbox{}+\int_{-\infty}^{t}e^{\varepsilon(t'-t)}\Bigl\{
\frac{i\vec{k}}{m}\cdot W_{n\dot{h}}(\vec{k};\vec{p};t,t')
-\frac{\partial}{\partial \vec{p}}\cdot
W_{Fh}(\vec{k};\vec{p};t,t')\nonumber
\\&&\mbox{}-\frac{\beta}{m}f_0(p)\vec{p}\cdot
W_{F\dot{h}}(\vec{k};\vec{p};t,t')\Bigr\}h_{\vec{k}}^{int}(t')dt',\nonumber
\end{eqnarray}
\begin{eqnarray}
\label{math/3.28} \lefteqn{\frac{\partial}{\partial
t}h_{\vec{k}}^{int}(t)=\int d\vec{p}'i\Omega_{hn}(\vec{k;\vec{p}'})f_{\vec{k}}(\vec{p}';t)}
\\&&\mbox{}-\int_{-\infty}^{t}e^{\varepsilon(t'-t)}
\Bigl\{\bar{\varphi}^{(0)}_{\dot{h}\dot{h}}(\vec{k};t,t')
-\vec{\Phi}'_{hF}(\vec{k})\cdot \bar{\varphi}^{(0)}_{\jmath
h}(\vec{k};t,t')\nonumber
\\&&\mbox{}-\bar{\varphi}_{h\jmath}(\vec{k};t,t')\cdot\vec{\Phi}'_{hF}(\vec{k})
+\vec{\Phi}'_{hF}(\vec{k})\cdot
\bar{\varphi}^{(0)}_{\jmath\jmath}(\vec{k};t,t')\cdot\vec{\Phi}_{Fh}(\vec{k})
\Bigr\}h_{\vec{k}}^{int}(t')dt' \nonumber
\\&&\mbox{}-\frac{i\vec{k}}{m}\cdot\int d\vec{p}'
\int_{-\infty}^{t}e^{\varepsilon(t'-t)}
W_{hn}(\vec{k};\vec{p}';t,t')\frac{1}{nf_0(p')}f_{\vec{k}}(\vec{p}';t')dt'
\nonumber
\\&&\mbox{}+\int d\vec{p}'\int_{-\infty}^{t}e^{\varepsilon(t'-t)}
W_{hF}(\vec{k};\vec{p}';t,t')\frac{1}{nf_0(p')}\cdot
\left(\frac{\beta}{mn}\vec{p}'-\frac{\partial}{\partial
\vec{p}'}\right)f_{\vec{k}}(\vec{p}';t')dt' \nonumber
\\&&\mbox{}-\int_{-\infty}^{t}e^{\varepsilon(t'-t)}
W_{h\jmath}(\vec{k};t,t')\frac{\beta}{mn}\cdot
\vec{\jmath}_{\vec{k}}(t')dt', \nonumber
\end{eqnarray}
where in the first equation the transport kernels have the
following structure:
\begin{eqnarray*}
\lefteqn{W_{n\dot{h}}(\vec{k};\vec{p};t,t')
=\bar{\varphi}_{\jmath\dot{h}}(\vec{k};\vec{p};t,t')}
\\&&\mbox{}-\vec{p}\cdot\bar{\varphi}_{n\dot{h}}(\vec{k};\vec{p};t,t')
-{\varphi}_{\jmath\jmath}(\vec{k};\vec{p};t,t')\cdot\vec{\Phi}'_{hF}(\vec{k})
+\vec{p}\cdot\varphi_{n\jmath}(\vec{k};\vec{p};t,t')\cdot\vec{\Phi}'_{hF}(\vec{k}),
\end{eqnarray*}
\begin{eqnarray*}
W_{Fh}(\vec{k};\vec{p};t,t')
=\bar{\varphi}_{Fh}(\vec{k};\vec{p};t,t') -{\varphi}_{F
\jmath}(\vec{k};\vec{p};t,t')\cdot\vec{\Phi}'_{hF}(\vec{k}) ,
\end{eqnarray*}
\begin{eqnarray*}
W_{F\dot{h}}(\vec{k};\vec{p};t,t')
=\vec{\Phi}_{Fh}(\vec{k})\cdot{\varphi}_{h\dot{h}}(\vec{k};t,t')
-\vec{\Phi}_{Fh}(\vec{k})\cdot{\varphi}_{h
\jmath}(\vec{k};t,t')\cdot\vec{\Phi}'_{hF}(\vec{k}).
\end{eqnarray*}

The transport equations (\ref{math/3.27}), (\ref{math/3.28}) are
functionally connected concerning the basic parameters of reduced
description $f_{\vec{k}}(\vec{p};t)$, $h^{int}_{\vec{k}}(t)$.
However, the equations contain the average values of densities of
particles number $\langle\hat{n}_{\vec{k}}\rangle^t$ and momentum
$\langle\hat{\vec{\jmath}}_{\vec{k}}\rangle^t$, which, generally
speaking, are the hydrodynamic variables. Integrating the equation
(\ref{math/3.27}) over momentum, we obtain the equation for
$n_{\vec{k}}(t)$
\begin{eqnarray}
\label{math/3.29} \frac{\partial}{\partial
t}n_{\vec{k}}(t)+\frac{i\vec{k}}{m} \cdot \vec{\jmath}_{\vec{k}}(t)=0,
\end{eqnarray}
that represents
the conservation law for average value of number of particles.

It is worth noting that the generalized diffusion and friction coefficients (\ref{math/3.9})--(\ref{math/3.15}) in the phase space together with
the generalized coefficients (\ref{math/3.24})--(\ref{math/3.25}), the diffusion coefficient
$\varphi^{(0)}_{\jmath\jmath}(\vec{k};t,t')$, the potential part of the generalized heat conductivity coefficient $\bar{\varphi}_{\dot{h}\dot{h}}^{(0)}(\vec{k};t,t')$ as well as together with coefficients $\bar{\varphi}_{\jmath\dot{h}}^{(0)}(\vec{k};t,t')$ and $\bar{\varphi}_{\dot{h}\jmath}(\vec{k};t,t')$ describing cross-correlations between viscous and heat processes all enter the set of transport equations (\ref{math/3.27}), (\ref{math/3.28}).

In contrast by the equations of molecular hydrodynamics \cite{77,Mryglod}, in this approach, the viscosity processes in the system are described by means of the generalized coefficients of diffusion and friction of particles in the phase space. Such a level of description is very important, in particular, in the case of molecular liquids, molecules of which possess their own structure and are sensitive to the process of momentum transfer
when interacting. Another significant feature of these equations is connected with the averaged value of the potential part of the enthalpy density $h_{\vec{k}}^{int}(t)$.
In the case when Fourier transform of the long-range part of the potential of interaction $\Phi(|\vec{r}_{lj}|)$ exists, the potential part of the energy density $\hat{\varepsilon}_{\vec{k}}^{int}$ can be expressed via the Fourier-components of particles number density:
\begin{equation}
\label{math/3.291}
\hat{\varepsilon}_{\vec{k}}^{int}=\frac{1}{2}\sum_{\vec{q}}\nu (q)
\hat{n}_{\vec{q}+\vec{k}}\hat{n}_{-\vec{q}},
\end{equation}
where $\nu (q)$ is the Fourier transform of the pair potential of interaction.
Taking into account (\ref{math/3.291}) the averaged value of the potential part of the enthalpy density can be presented as follows:
\begin{equation}
\label{math/3.292}
h_{\vec{k}}^{int}(t)=\frac{1}{2}\sum_{\vec{q}}\nu (q)F_{2}(\vec{q}+\vec{k},-\vec{q};t)+\frac{1}{2}\sum_{\vec{q}}\nu (q)S_{3}(\vec{q}+\vec{k},-\vec{q},-\vec{k})S_{2}^{-1}(\vec{q})n_{\vec{k}}(t).
\end{equation}
Here, $F_{2}(\vec{q}+\vec{k},-\vec{q};t)=\langle \hat{n}_{\vec{q}+\vec{k}}\hat{n}_{-\vec{q}}\rangle^{t}$ is the nonequilibrium scattering function, whose Fourier transform is the nonequilibrium dynamic structure factor $S_{2}(\vec{q}+\vec{k},-\vec{q};\omega)$ of particles of the system. $S_{3}(\vec{q}+\vec{k},-\vec{q},-\vec{k})=\langle \hat{n}_{\vec{q}+\vec{k}}\hat{n}_{-\vec{q}}\hat{n}_{-\vec{k}}\rangle_{0}$ is the three-particles equilibrium structure factor. The function $S_{2}(\vec{q}+\vec{k},-\vec{q};t)$ is important from the point of view of describing the dynamics of neutron scattering  in the system \cite{Temper,74,Bul}.

On the basis of the generalized transport equations (\ref{math/3.27}), (\ref{math/3.28}) one can obtain the corresponding set of equations for time correlation functions~\cite{cmp2010}.

Let us now project the set of equations (\ref{math/2.11}), (\ref{math/2.12}) or (\ref{math/3.27}),
(\ref{math/3.28}) onto the first moments of the nonequilibrium
one-particle distribution function
\begin{eqnarray}
\Psi_1(\vec{p})=1, \quad
\Psi_{\alpha}(\vec{p})=\sqrt{2}p_{\alpha}/2k_BT, \quad
\Psi_{\varepsilon}(\vec{p})=\sqrt{2/3}(p^2/2mk_BT-3/2), \quad
(\alpha=x,y,z).\nonumber\\
\end{eqnarray}
Then, we obtain the set of equations for the averaged values of densities
of particles number $n_{\vec{k}}(t)$, momentum
$\vec{\jmath}_{\vec{k}}(t)$, kinetic $h_{\vec{k}}^{kin}(t)$ and
potential $h_{\vec{k}}^{int}(t)$ parts of enthalpy~\cite{tok}, where the Fourier-components of the kinetic part of enthalpy density defined as $\hat{h}_{\vec{k}}^{kin}=\hat{\varepsilon}_{\vec{k}}^{kin}
-\langle\hat{\varepsilon}_{\vec{k}}^{kin}\hat{n}_{-\vec{k}}\rangle_0
\langle\hat{n}_{\vec{k}}\hat{n}_{-\vec{k}}\rangle_0^{-1}\hat{n}_{\vec{k}}$.
Using the Laplace transform let us represent the obtained system of equations for averages $\tilde{a}_{\vec{k}}(z)=\{n_{\vec{k}}(z),
\vec{\jmath}_{\vec{k}}(z), h_{\vec{k}}^{kin}(z),
h_{\vec{k}}^{int}(z)\}$ in a matrix form:
\begin{eqnarray}
\label{math/3.432}
z\tilde{a}_{\vec{k}}(z)-\tilde{\Sigma}^G(\vec{k};z)\tilde{a}_{\vec{k}}(z)=-\langle
\tilde{a}_{\vec{k}}(t=0)\rangle^{t}.
\end{eqnarray}
with the matrix $\tilde{\Sigma}^{G}(\vec{k};z)$.
\begin{eqnarray}
\label{math/3.44} \tilde{\Sigma}^G(\vec{k};z) =i\tilde\Omega^G(\vec{k})+\tilde\Pi^G(\vec{k};z),
\end{eqnarray}
where
\begin{eqnarray}
\label{math/3.45} i\tilde{\Omega}^G(\vec{k})=\left(
\begin{array}{llll}
  0       & i\Omega_{n\jmath} & 0 & 0 \\
  i\Omega_{\jmath n} & 0 & i\Omega_{\jmath h}^{kin} & i\Omega_{\jmath h}^{int} \\
  0 & i\Omega_{h\jmath}^{kin} & 0 & 0 \\
  0 & i\Omega_{h\jmath}^{int} & 0 & 0 \\
\end{array}\right)_{(\vec{k})}
\end{eqnarray}
is the frequency matrix, and
%
%
\begin{eqnarray}
\label{math/3.46} \tilde{\Pi}^G(\vec{k};z)=\left(
\begin{array}{llll}
  0 & 0 & 0 & 0 \\
  0 & \Pi_{\jmath\jmath} & \Pi_{\jmath h}^{kin} & \Pi_{\jmath h}^{int} \\
  0 & \Pi_{h\jmath}^{kin} & \Pi_{hh}^{kin,kin} & \Pi_{hh}^{kin,int} \\
  0 & \Pi_{h\jmath}^{int} & \Pi_{hh}^{int,kin} & \Pi_{hh}^{int,int} \\
\end{array}\right)_{(\vec{k};z)}
\end{eqnarray}
is the matrix of transport kernels. Its elements have the
following structure:
\begin{eqnarray}
\label{math/3.47} \Pi_{\mu\nu}(\vec{k};z)
=\langle\Psi_{\mu}|\tilde{\varphi}(\vec{k};z)+\tilde{\Sigma}(\vec{k};z)
{\cal Q}[z\tilde{I}-{\cal Q}\tilde{\Sigma}(\vec{k};z){\cal
Q}]^{-1}{\cal Q}\tilde{\Sigma}(\vec{k};z)|\Psi_{\nu}\rangle,
\end{eqnarray}
where ${\cal Q}=1-{\cal P}$ with ${\cal P}$ being the projection operator
constructed on the first moments $|\Psi_{\alpha}(\vec{p})\rangle$
of the nonequilibrium one-particle distribution function. This operator acts according to the rule ${\cal
P}\langle\Psi|=\sum_{\nu=1}^n\langle\Psi|\Psi_{\nu}\rangle\langle\Psi_{\nu}|$, where
$\langle\Psi|\Psi_{\nu}\rangle=\int
d\vec{\xi}\Psi(\vec{\xi})f_0(\xi)\Psi_{\nu}(\vec{\xi})$ is the scalar product of two functions.
$\{\Psi_{\nu}(\xi)\}$ is a set of functions satisfying the conditions
$\langle\Psi_{\mu}|\Psi_{\nu}\rangle=\delta_{\mu\nu}$,
$\sum_{\nu}|\Psi_{\nu}\rangle\langle\Psi_{\nu}|=1$. As we can see
from the structure of elements of the matrixes $i\tilde\Omega^G(\vec{k})$ and $\tilde\Pi^G(\vec{k};z)$,
the contributions of kinetic and potential parts of enthalpy are
separated. However, all the transport kernels of
$\tilde\Pi(\vec{k};z)$ are determined in terms of the time
correlation functions (\ref{math/3.17}), (\ref{math/3.19}) and the
transport kernels
$\bar{\varphi}^{(0)}_{\dot{h}\dot{h}}(\vec{k};t,t')$,
$\bar{\varphi}^{(0)}_{\jmath\dot{h}}(\vec{k};t,t')$,
$\bar{\varphi}^{(0)}_{\dot{h}\jmath}(\vec{k};t,t')$ and
$D(\vec{k};t,t')$ (\ref{math/3.25}).

\section{Transition to equations of molecular hydrodynamics}\label{sec:4}

It is well known that the molecular hydrodynamics of dense gases and liquids is based on the transport equations for the averaged values of densities of particles number $\langle\hat{n}_{\vec k}\rangle^t$, momentum $\langle\hat{\vec\jmath}_{\vec k}\rangle^t$ and generalized enthalpy $\langle\hat{h}_{\vec k}\rangle^t$~\cite{77,MrygTok}. In our case $\langle\hat{h}_{\vec k}\rangle^t=\langle\hat{h}_{\vec k}^{kin}\rangle^t+\langle\hat{h}_{\vec k}^{int}\rangle^t$.
As a result of projecting of the equations (\ref{math/3.27}), (\ref{math/3.28}) onto the first moments of the nonequilibrium one-particle distribution function we obtain a set of equations for the averages $\langle\hat{n}_{\vec k}\rangle^t$, $\langle\hat{\vec\jmath}_{\vec k}\rangle^t$, $\langle\hat{h}_{\vec k}^{kin}\rangle^t$ and $\langle\hat{h}_{\vec k}^{int}\rangle^t$
[with the elements of frequency matrix (\ref{math/3.45}) and matrix of memory functions (\ref{math/3.46})], which in the Laplace representation we write down in the explicit form:
\begin{eqnarray}
\label{math/4.1}
&& zn_{{\vec k}}(z)+i\Omega_{n\jmath}({\vec k})\vec{\jmath}_{{\vec k}}(z)=-\langle\hat{n}_{\vec k} (t=0)\rangle,\\
&& \label{math/4.2}
z\vec{\jmath}_{\vec k}(z)\,+i\Omega_{\jmath n}({\vec k})n_{\vec k}(z)+i\Omega_{jh}^{kin}({\vec k})h^{kin}_{\vec k}(z)
+i\Omega_{\jmath h}^{int}({\vec k})h^{int}_{\vec k}(z)\nonumber\\
&&\hspace{10mm}\,-\Pi_{\jmath\jmath}({\vec k},z)\vec{\jmath}_{\vec k}(z)
-\Pi_{\jmath h}^{kin}({\vec k},z)h_{\vec k}^{kin}(z)-\Pi_{\jmath h}^{int}({\vec k},z)h_{\vec k}^{int}(z)=-\langle\hat{\vec\jmath}_{\vec k} (t=0)\rangle,\\
&& \label{math/4.3}
zh_{\vec k}^{kin}(z)+i\bar{\Omega}_{h\jmath}^{kin}({\vec k},z)\vec{\jmath}_{\vec k}(z)\nonumber\\
&&\hspace{10mm}\,-\Pi_{hh}^{kin,kin}({\vec k},z)h^{kin}_{\vec k}(z)
-\Pi_{hh}^{kin,int}({\vec k},z)h^{int}_{\vec k}(z)=-\langle\hat{h}^{kin}_{\vec k} (t=0)\rangle,\\
&& \label{math/4.4}
zh_{\vec k}^{int}(z)+i\bar{\Omega}_{h\jmath}^{int}({\vec k},z)\vec{\jmath}_{\vec k}(z)\nonumber \\
&&\hspace{10mm}\,-\Pi_{hh}^{int,kin}({\vec k},z)h^{kin}_{\vec k}(z)
-\Pi_{hh}^{int,int}({\vec k},z)h^{int}_{\vec k}(z)=-\langle\hat{h}^{int}_{\vec k} (t=0)\rangle,
\end{eqnarray}
where
\begin{eqnarray}
\label{math/4.5}
&&i\bar{\Omega}_{h\jmath}^{kin}({\vec k},z)=i{\Omega}_{h\jmath}^{kin}({\vec k})-\Pi_{h\jmath}^{kin}({\vec k},z),\nonumber\\
&&i\bar{\Omega}_{h\jmath}^{int}({\vec k},z)=i{\Omega}_{h\jmath}^{int}({\vec k})-\Pi_{h\jmath}^{int}({\vec k},z).
\end{eqnarray}
In order to pass to the equations of  molecular hydrodynamics let us sum up equations
(\ref{math/4.3}) and (\ref{math/4.4}). As a result we obtain (superscript ``tot'' denotes the total enthalpy density):
\begin{eqnarray}
\label{math/4.6}
&&zh_{{\vec k}}(z)+i\bar{\Omega}_{h\jmath}({\vec k},z)\vec{\jmath}_{\vec k}(z)
-\Pi_{hh}^{tot,kin}({\vec k},z)h_{\vec k}^{kin}(z)-\Pi_{hh}^{tot,int}({\vec k},z)h_{\vec k}^{int}(z)
=\langle\hat{h}_{\vec k}(t=0)\rangle,
\end{eqnarray}
where
\begin{eqnarray}
\label{math/4.7}
h_{\vec k}(z)&=&h_{\vec k}^{kin}(z)+h_{\vec k}^{int}(z),\\
\label{math/4.8}
i\bar{\Omega}_{h\jmath}({\vec k},z)&=&i{\Omega}_{h\jmath}({\vec k})-\Pi_{h\jmath}({\vec k},z).
\end{eqnarray}
Then, after some transformations we can preset this equation in the following form:
\begin{align}\label{math/4.9}
zh_{\vec k}(z)&+i\bar{\Omega}_{h\jmath}({\vec k},z)\vec{\jmath}_{\vec k}(z)
-\Pi_{hh}({\vec k},z)h_{\vec k}(z)\nonumber\\
&+\Pi_{hh}^{tot,kin}({\vec k},z)h_{\vec k}^{int}(z)+\Pi_{hh}^{tot,int}({\vec k},z)h_{\vec k}^{kin}(z)
=-\langle\hat{h}_{\vec k}(t=0)\rangle,
\end{align}
where
\begin{eqnarray}\label{math/4.10}
\Pi_{hh}({\vec k},z)=\Pi_{hh}^{tot,kin}({\vec k},z)+\Pi_{hh}^{tot,int}({\vec k},z).
\end{eqnarray}
To eliminate  $h_{\vec k}^{int}(z)$ and $h_{\vec k}^{kin}(z)$ from (\ref{math/4.9}) we use equations (\ref{math/4.3}), (\ref{math/4.4}) once more, and find:
\begin{align}
\label{math/4.11}
zh_{\vec k}^{int}(z)&+i\tilde{\Omega}_{h\jmath}^{int}({\vec k},z)\vec{\jmath}_{\vec k}(z)
-\Sigma_{hh}^{int,int}({\vec k},z)h_{\vec k}^{int}(z)=0,\\\nonumber
%
\label{math/4.12}
zh_{\vec k}^{kin}(z)&+\Big\{
i\bar{\Omega}_{h\jmath}^{kin}({\vec k},z)+\Pi_{hh}^{kin,int}({\vec k},z)
\left[z-\Sigma_{hh}^{int,int}({\vec k},z)\right]^{-1}i\tilde{\Omega}_{h\jmath}^{int}({\vec k},z)\Big\}\vec{\jmath}_{\vec k}(z)\nonumber\\
&-\Pi_{hh}^{kin,kin}({\vec k},z)h_{\vec k}^{kin}(z)=0,
\end{align}
where
\begin{eqnarray}
\label{math/4.13}
i\tilde{\Omega}_{h\jmath}^{int}({\vec k},z)&=&i\bar{\Omega}_{h\jmath}^{int}({\vec k},z)
+\Pi_{hh}^{int,kin}({\vec k},z)\left[z-\Pi_{hh}^{kin,kin}({\vec k},z)\right]^{-1} i\bar{\Omega}_{h\jmath}^{kin}({\vec k},z),\\
\label{math/4.14}
\Sigma_{hh}^{int,int}({\vec k},z)&=&\Pi_{hh}^{int,int}({\vec k},z)
+\Pi_{hh}^{int,kin}({\vec k},z)\left[z-\Pi_{hh}^{kin,kin}({\vec k},z)\right]^{-1}\Pi_{hh}^{kin,int}({\vec k},z).
\end{eqnarray}
After expressing  $h_{\vec k}^{int}(z)$ and $h_{\vec k}^{kin}(z)$ from (\ref{math/4.11}), (\ref{math/4.12}) we can present equation (\ref{math/4.9}) as
\begin{eqnarray}
\label{math/4.15}
&&zh_{\vec k}(z)+\left(i\bar{\Omega}_{h\jmath}({\vec k},z)-\Sigma_{h\jmath}({\vec k},z)\right)\vec{\jmath}_{\vec k}(z)
-\Pi_{hh}({\vec k},z)h_{\vec k}(z)=-\langle\hat{h}_{\vec k}(t=0)\rangle
\end{eqnarray}
with
\begin{eqnarray}
\label{math/4.16}
\Sigma_{h\jmath}({\vec k},z)&=&\Pi_{hh}^{tot,kin}({\vec k},z)\frac{1}{z-\Sigma_{hh}^{int,int}({\vec k},z)}
i\bar{\Omega}_{h\jmath}^{int}+\Pi_{hh}^{tot,int}({\vec k},z)\frac{1}{z-\Pi_{hh}^{kin,kin}({\vec k},z)}\nonumber\\
&&\times\left(i\bar{\Omega}_{h\jmath}^{kin}({\vec k},z)+\Pi_{hh}^{kin,int}({\vec k},z)\frac{1}{z-\Sigma_{hh}^{int,int}({\vec k},z)}
i\tilde{\Omega}_{h\jmath}^{int}({\vec k},z)\right).
\end{eqnarray}

Now let us write down equation (\ref{math/4.2}) in the following form:
\begin{align}
\label{math/4.17}
z\vec{\jmath}_{\vec k}(z) & +i{\Omega}_{\jmath n}({\vec k})n_{\vec k}(z)+i\bar{\Omega}_{\jmath h}({\vec k},z)h_{\vec k}(z)
-\Pi_{\jmath\jmath}({\vec k},z)\vec{\jmath}_{\vec k}(z)\nonumber\\
&-i\bar{\Omega}_{\jmath h}^{int}({\vec k},z)h^{kin}_{\vec k}(z)
-i\bar{\Omega}_{\jmath h}^{int}({\vec k},z)h^{int}_{\vec k}(z)
=-\langle\hat{\vec{\jmath}}_{\vec k}(t=0)\rangle,
\end{align}
where
\begin{eqnarray}
\label{math/4.18}
i\bar{\Omega}_{\jmath h}({\vec k},z)h_{\vec k}(z)&=&i{\Omega}_{\jmath h}({\vec k})
-\Pi_{\jmath h}({\vec k},z),\\
\label{math/4.19}
i\bar{\Omega}_{\jmath h}^{int}({\vec k},z)h_{\vec k}(z)&=&i{\Omega}_{\jmath h}^{int}({\vec k})
-\Pi_{\jmath h}^{int}({\vec k},z),\\
\label{math/4.20}
i\bar{\Omega}_{\jmath h}^{kin}({\vec k},z)h_{\vec k}(z)&=&i{\Omega}_{\jmath h}^{kin}({\vec k})
-\Pi_{\jmath h}^{kin}({\vec k},z).
\end{eqnarray}

After taking into account  (\ref{math/4.11}) and (\ref{math/4.12}), we can present equation (\ref{math/4.9}) in the form
\begin{eqnarray}
\label{math/4.21}
&&z\vec{\jmath}_{\vec k}(z)+i{\Omega}_{\jmath n}({\vec k})n_{\vec k}(z)+i\bar{\Omega}_{\jmath h}({\vec k},z)h_{\vec k}(z)
-\Sigma_{\jmath\jmath}({\vec k},z)\vec{\jmath}_{\vec k}(z)=-\langle\hat{\vec{\jmath}}_{\vec k}(t=0)\rangle
\end{eqnarray}
with
\begin{eqnarray}
\label{math/4.22}
\Sigma_{\jmath\jmath}({\vec k},z)&=&\Pi_{\jmath\jmath}({\vec k},z)-i\bar{\Omega}_{\jmath h}^{int}({\vec k},z)
\frac{1}{z-\Pi_{hh}^{kin,kin}({\vec k},z)}\nonumber\\
&&\times\left\{i\bar{\Omega}_{h\jmath}^{kin}({\vec k},z)
+\Pi_{hh}^{kin,kin}({\vec k},z)\frac{1}{z-\Sigma_{hh}^{int,int}({\vec k},z)}i\tilde{\Omega}_{h\jmath}^{kin}({\vec k},z)\right\}
\nonumber\\
&&-i\bar{\Omega}_{\jmath h}^{kin}({\vec k},z)\frac{1}{z-\Sigma_{hh}^{int,int}({\vec k},z)}i\tilde{\Omega}_{h\jmath}^{int}({\vec k},z).
\end{eqnarray}

Finally, the system of equation, which by its structure coincide with the set of equations of molecular hydrodynamics, can be presented as follows:
\begin{align}
\label{math/4.23}
zn_{\vec k}(z)&+i{\Omega}_{n\jmath}({\vec k})\vec{\jmath}_{\vec k}(z)=-\langle\hat{n}_{\vec k}(t=0)\rangle,\\
\label{math/4.24}
z\vec{\jmath}_{\vec k}(z)\,&+i{\Omega}_{\jmath n}({\vec k})n_{\vec k}(z)+i{\Omega}_{\jmath h}({\vec k},z)h_{\vec k}(z)\nonumber\\
&-\Pi_{\jmath h}({\vec k},z)h_{\vec k}(z)
-\Sigma_{\jmath \jmath}({\vec k},z)\vec{\jmath}_{\vec k}(z)
=-\langle\hat{\vec{\jmath}}_{\vec k}(t=0)\rangle,\\
\label{math/4.25}
zh_{\vec k}(z)&+i{\Omega}_{h\jmath}({\vec k})\vec{\jmath}_{\vec k}(z)-\Pi_{h\jmath}({\vec k},z)\vec{\jmath}_{\vec k}(z)
-\Pi_{hh}({\vec k},z)h_{\vec k}(z)
=-\langle\hat{h}_{\vec k}(t=0)\rangle.
\end{align}
Here, $\Sigma_{\jmath \jmath}({\vec k},z)$,  $\Pi_{hh}({\vec k},z)$, $\Pi_{h\jmath}({\vec k},z)$ and  $\Pi_{\jmath h}({\vec k},z)$ determine the generalized viscosity, heat conductivity coefficients as well as the generalized transport cross-coefficients describing dissipative correlations between viscous and heat processes.

It is worth to mention that the presented generalized transport kernels
are expressed via the transport kernels
$\varphi_{nn}(\vec{k};\vec{p},\vec{p}';t,t')$,  $\varphi_{nh}(\vec{k};\vec{p};t,t')$, $\varphi_{hn}(\vec{k};\vec{p}';t,t')$ and $\varphi_{hh}(\vec{k};t,t')$ of a consistent description of
kinetics and hydrodynamics and, obviously, their calculation depends on an explicit model of interparticle interaction.

\section{Spectrum of collective excitations}\label{sec:5}

Let us consider the system of kinetic equations~(\ref{math/2.11}),
(\ref{math/2.12}) in the case when the potential of interaction is
presented as follows:
\begin{eqnarray}
\label{math/3.48} \Phi(|\vec{r}_{ij}|)=\Phi^\mathrm{hs}(|\vec{r}_{ij}|)
+\Phi^\mathrm{l}(|\vec{r}_{ij}|),
\end{eqnarray}
where $\Phi^\mathrm{hs}(|\vec{r}_{ij}|)$ is the hard sphere interaction
potential, and $\Phi^\mathrm{l}(|\vec{r}_{ij}|)$ is the long-range
potential. Taking into account the features of the hard sphere
model dynamics~\cite{zub6} and the results of investigations
\cite{117,114,deSchep}, one can separate Enskog-Boltzmann
collision integral from the function
$\varphi_{nn}(\vec{k};\vec{p},\vec{p}';t,t')$. Indeed, an
infinitesimal time of a collision $\tau_0\rightarrow+0$ within an
infinitesimal region $\sigma\pm\Delta r_0$, $\Delta
r_0\sim|\tau_0||\vec{p}_2-\vec{p}_1|/m\rightarrow+0$ being a
feature of the hard sphere model dynamics ($\sigma$ is the
hard sphere diameter). Taking this into account in the kinetic
equation~(\ref{math/2.11}) we can obtain the kinetic equation of the revised Enskog
theory for the hard sphere model and the kinetic Enskog-Landau
equation for the charged hard sphere model in a pair collision
approximation, respectively~\cite{zub6}. In the latter case, when
$\Phi^\mathrm{l}(|\vec{r}_{ij}|)$ is the Coulomb potential of interaction,
taking into account the features $\tau_0\rightarrow+0$, $\Delta
r_0\rightarrow+0$ makes it possible to separate a collision
integral of the revised Enskog theory and a Landau-like collision
integral in the limits $\tau\rightarrow-0$ and
$\tau\rightarrow-\infty$, respectively. In the case of potential
(\ref{math/3.48}), in the region of $\tau_0\rightarrow+0$, $\Delta
r_0\rightarrow+0$, $\sigma\pm\Delta r_0$ when the main
contribution to a dynamics is defined by pair collisions of hard
spheres, the memory function
$\varphi_{nn}(\vec{k};\vec{p},\vec{p}';t,t')$ can be calculated by
expanding it over the density (a pair collision approximation),
what was done in papers by Mazenko~\cite{109,110,111,114,117} in
detail.

Then, the kinetic equation~(\ref{math/2.11}) can be represented in
the following form:
\begin{eqnarray}
\label{math/3.49} \lefteqn{\frac{\partial}{\partial
t}f_{\vec{k}}(\vec{p};t)+\frac{i\vec{k}\cdot\vec{p}}{m}f_{\vec{k}}(\vec{p};t)
=-\frac{i\vec{k}\cdot\vec {p}}{m}nf_{0}(\vec{p})c_2(k)\int
d\vec{p}'f_{\vec{k}}(\vec{p}';t)}\nonumber
\\&&\mbox{}-\int
d\vec{p}'
\varphi_{nn}^\mathrm{hs}(\vec{k},\vec{p},\vec{p}')f_{\vec{k}}(\vec{p}';t)
+i\Omega_{nh}(\vec{k};\vec{p})h_{\vec{k}}^\mathrm{int}(t) \nonumber
\\&&\mbox{}-\int
d\vec{p}'\int_{-\infty}^t d t'e^{\varepsilon(t-t')}
\varphi_{nn}^\mathrm{l}(\vec{k};\vec{p},\vec{p}';t,t')f_{\vec{k}}(\vec{p}';t')
-\int_{-\infty}^t d t'e^{\varepsilon(t-t')}\varphi_{nh}(\vec{k};\vec{p};t,t')h_{\vec{k}}^\mathrm{int}(t').\nonumber\\
\end{eqnarray}
Here,
\begin{eqnarray}
\label{math/3.49a} \lefteqn{\int d\vec{p}'
\varphi_{nn}^\mathrm{hs}(\vec{k},\vec{p},\vec{p}')f_{\vec{k}}(\vec{p}';t)=
 ng_2(\sigma)\sigma^2\int d\Omega_{\sigma}\int
d\vec{p}'\frac{(\vec{p}-\vec{p}')\cdot\hat{\vec{\sigma}}}{m}
\Theta_{-}\left(\hat{\vec{\sigma}}\cdot[\vec{p}-\vec{p}']\right)}\nonumber
\\&&\mbox{}
\times\left[f_0(p'^*)f_{\vec{k}}(\vec{p};t)-f_0(p')f_{\vec{k}}(\vec{p}^*;t) +
e^{i\vec{k}\cdot\hat{\vec{\sigma}}\sigma}f_0(p'^*)f_{\vec{k}}(\vec{p}'^*;t)-
e^{i\vec{k}\cdot\hat{\vec{\sigma}}\sigma}f_0(p)f_{\vec{k}}(\vec{p}';t)\right]
\end{eqnarray}
is the Enskog-Boltzmann collision integral, where
$c_2^0(\vec{k})$ is the low-density limit of the direct
correlation function and $g_2(\sigma)$ is the pair distribution
function. The step function $\Theta_-(x)$ is unity for $x<0$ and
vanishes otherwise. $d\Omega_\sigma$ is the differential solid
angle, $\hat{\vec{\sigma}}$ is unity vector. The pre- and
postcollision momenta of the colliding hard spheres are denoted as
$(\vec{p},\vec{p}')$ and $(\vec{p}^*,\vec{p}'^*)$, respectively.
$\varphi_{nn}^\mathrm{l}(\vec{k};\vec{p},\vec{p}';t,t')$ is the part of the
transport kernel related to the long-range interaction potential
$\Phi^\mathrm{l}(|\vec{r}_{ij}|)$. Notably, the presented equation contains
the Enskog-Boltzmann collision integral describing short-time
dynamics of the hard sphere model. The collective effects related
to the long-range interactions between particles are described by
the functions $i\Omega_{nh}(\vec{k};\vec{p})$,
$\varphi_{nn}^\mathrm{l}(\vec{k};\vec{p},\vec{p}';t,t')$,
$\varphi_{nh}(\vec{k};t,t')$ and by the equation for
$h_{\vec{k}}^\mathrm{int}(t)$. Since the collective modes for the
Enskog-Boltzmann model are well studied~\cite{deSchep}, the
investigation of time correlation functions and collective modes
for the system of particles interacting through the
potential~(\ref{math/3.48}) turns out to be of great interest. In
the case of the hard sphere system, the set of kinetic
equations~(\ref{math/2.12}), (\ref{math/3.49}) reduces to the
Enskog-Boltzmann kinetic equation~\cite{deSchep}.
\begin{eqnarray}
\label{math/3.50} \lefteqn{\frac{\partial}{\partial
t}f_{\vec{k}}(\vec{p};t)+\frac{i\vec{k}\cdot\vec{p}}{m}f_{\vec{k}}(\vec{p};t)=-\frac{i\vec{k}\cdot\vec{p}}{m}nf_{0}(\vec{p})\left[c_2(k)-g_2(\sigma)c_2^0(k)\right]\int
d\vec{p}'f_{\vec{k}}(\vec{p}';t)}\nonumber
\\&&\mbox{}-ng_2(\sigma)\sigma^2\int
d\Omega_{\sigma}\int
d\vec{p}'\frac{(\vec{p}-\vec{p}')\cdot\hat{\vec{\sigma}}}{m}
\Theta_{-}\left(\hat{\vec{\sigma}}\cdot[\vec{p}-\vec{p}']\right)\nonumber
\\&&\mbox{}\times\left[
f_0(p'^*)f_{\vec{k}}(\vec{p};t)-f_0(p')f_{\vec{k}}(\vec{p}^*;t)+
e^{i\vec{k}\cdot\hat{\vec{\sigma}}\sigma}f_0(p'^*)f_{\vec{k}}(\vec{p}'^*;t)-
e^{i\vec{k}\cdot\hat{\vec{\sigma}}\sigma}f_0(p)f_{\vec{k}}(\vec{p}';t)\right].
\end{eqnarray}
Projecting the Enskog-Boltzmann equation onto the first moments of the
nonequilibrium one-particle distribution function a spectrum of
collective excitations for the hard sphere model was obtained
in~\cite{deSchep,deSchep2}.  Herewith, it is important to note
that for the kinetic Enskog-Boltzmann equation we can consider two
typical limits: $k\sigma\ll1$ and $k\sigma\gg1$. In the case when
$k\sigma\ll1$ the spectrum includes:
\emph{heat mode} $z_\mathrm{H}(k)=-D_\mathrm{TE}k^2$, where $D_\mathrm{TE}$ is
    the thermal diffusivity coefficient in the
    Enskog transport theory~\cite{Chap};
\emph{two sound modes} with eigenvalues given by
    $z_{\pm}(k)=\pm i ck-\Gamma_\mathrm{E}k^2$, where
    $\Gamma_\mathrm{E}$ is the sound damping coefficient and $c$ is the sound velocity
    in the Enskog theory;
\emph{two transverse modes} with eigenvalues
    given by $z_{\nu_1}(k)=z_{\nu_2}(k)=z_{\nu}(k)=-\nu_\mathrm{E}k^2$,
    $\nu_\mathrm{E}$ is the kinematic viscosity in the Enskog dense gas
    theory.
In the limit when $k\sigma\gg1$ the Enskog-Boltzmann collision
integral (\ref{math/3.49a}) is transformed~\cite{deSchep} into the
Lorentz-Boltzmann collision integral which has only one
eigenfunction $\Psi_1(\vec{p})=1$. Consequently, we obtain the
\emph{diffusion mode} only with the eigenvalue
    $z_\mathrm{D}(k)=-D_\mathrm{E}k^2$, $D_\mathrm{E}$ is the self-diffusion coefficient as given by the Enskog dense gas theory.

Let us now project the system of equation~(\ref{math/2.12}),
(\ref{math/3.49}) onto the first moments of the nonequilibrium
one-particle distribution function. Thereafter we perform a simple
transformations consisting in the transition from the set of
equations (\ref{math/3.432}) for averages $n_{\vec{k}}(z),
\vec{\jmath}_{\vec{k}}(z), h_{\vec{k}}^\mathrm{kin}(z),
h_{\vec{k}}^\mathrm{int}(z)$ to the equations of the generalized
hydrodynamics for averages
$\tilde{b}_{\vec{k}}(z)=[n_{\vec{k}}(z),
\vec{\jmath}_{\vec{k}}(z), h_{\vec{k}}(z)=h_{\vec{k}}^\mathrm{kin}(z)+
h_{\vec{k}}^\mathrm{int}(z)]$ as it was described in section~\ref{sec:4}. This permits to correctly define (see below) the generalized viscosity coefficient via the transport
kernel~(\ref{math/4.22}) and the heat conductivity coefficient via
the transport kernel $\Pi_{hh}(\vec{k},z)$. The averages
$\tilde{b}_{\vec{k}}(z)$ satisfy the set of equations
\begin{eqnarray}
z\tilde{b}_{\vec{k}}(z)-\tilde{\Sigma}^\mathrm{H}(\vec{k};z)\tilde{b}_{\vec{k}}(z)=-\langle
\tilde{b}_{\vec{k}}(t=0)\rangle^{t}.
\end{eqnarray}
In the limit $k\sigma\gg1$ the latter reduces to the single equation of diffusion for $n_{\vec{k}}(z)$ in which the transport kernel
$\Sigma^\mathrm{H}(\vec{k};z)=\langle\Psi_{1}|\varphi_{nn}^\mathrm{L-B}(\vec{k})|\Psi_{1}\rangle$ corresponds to the Lorentz-Boltzmann collision integral~(\ref{math/3.49a}). In the opposite case, when $k\sigma \ll 1$, matrix $\tilde{\Sigma}^\mathrm{H}(\vec{k};z)$ is defined as follows:
$\tilde{\Sigma}^\mathrm{H}(\vec{k};z)=i\tilde\Omega^\mathrm{H}(\vec{k})
-\tilde\Pi^\mathrm{H}(\vec{k};z)$,
\begin{eqnarray}
\label{math/3.466}
\tilde{\Sigma}_\mathrm{H}(\vec{k};z)=\left(
\begin{array}{lll}
   0 & i\Omega_{n\jmath} & 0 \\
   i\Omega_{\jmath n} & -\langle\Psi_{2}|\varphi_{nn}^\mathrm{hs}|\Psi_{2}\rangle-\Sigma^\mathrm{l}_{jj} &i\Omega_{\jmath h}-\Pi_{\jmath h} \\
  0 &i\Omega_{h\jmath}- \Pi_{h\jmath} & -\langle\Psi_{3}|\varphi_{nn}^\mathrm{hs}|\Psi_{3}\rangle-\Pi^\mathrm{l}_{hh} \\
  \end{array}\right)_{(\vec{k},z)}
\end{eqnarray}
with $\Sigma_{jj}(\vec{k},z)$ defined by (\ref{math/4.22}) and
\begin{eqnarray}
\label{math/3.61}
\Pi_{hh}(\vec{k},z)&=&\Pi_{hh}^\mathrm{kin,kin}(\vec{k},z)+\Pi_{hh}^\mathrm{kin,int}(\vec{k},z)
        +\Pi_{hh}^\mathrm{int,kin}(\vec{k},z)+\Pi_{hh}^\mathrm{int,int}(\vec{k},z),
\end{eqnarray}
which is equivalent to (\ref{math/4.10}).

We can separate real and imaginary parts in memory functions~(\ref{math/4.22}) and~(\ref{math/3.61}) as follows:
\begin{eqnarray}
\label{math/3.62}
\Sigma_{jj}(\vec{k},z)=\Sigma^{\prime}_{jj}(\vec{k},\omega)+i\Sigma^{\prime\prime}_{jj}(\vec{k},\omega),
\qquad
\Pi_{hh}(\vec{k},z)=\Pi^{\prime}_{hh}(\vec{k},\omega)+i\Pi^{\prime\prime}_{hh}(\vec{k},\omega).
\end{eqnarray}
Herewith, the contributions from the hard sphere dynamics with typical
spatial-temporal scale $\tau_0\rightarrow+0$, $\Delta
r_0\rightarrow+0$ are separated in the transport kernel
$\varphi_{nn}(\vec{k};\vec{p},\vec{p}';t,t')$ only in the first
term in the right-hand side of elements~(\ref{math/3.47}) and hence in~(\ref{math/4.22}). After these transformations we can obtain a spectrum of collective
excitations in the hydrodynamic limit $k\sigma\ll1$.
\emph{Heat mode} is defined as follows:
\begin{eqnarray}
\label{math/3.63}
z_\mathrm{H}(k)=-D_\mathrm{T}k^2, \qquad
D_\mathrm{T}=D_\mathrm{TE}+D_\mathrm{T}^\mathrm{l},
\end{eqnarray}
where $D_\mathrm{T}$ is the thermal diffusivity coefficient for the system with the
potential of interaction~(\ref{math/3.48}). In (\ref{math/3.63})
$D_\mathrm{T}^\mathrm{l}$ is determined through the corresponding
elements~(\ref{math/3.47}) of matrix of transport
kernels~(\ref{math/3.46}):
\begin{eqnarray}
\label{math/3.64}
D_\mathrm{T}^\mathrm{l}=\frac{\lambda^\mathrm{l}}{nmc_{p}}, \qquad
\lambda^\mathrm{l}=\lim_{\vec{k}\rightarrow
0,\omega\rightarrow 0}\lambda^\mathrm{l}(\vec{k},\omega), \qquad
\lambda^\mathrm{l}(\vec{k},\omega)=\frac{c_{v}(k)}{k_\mathrm{B}\beta^{2}}\frac{1}{k^{2}}
\Pi_{hh}^{\prime\prime}(\vec{k},\omega).
\end{eqnarray}
Here, $\lambda^\mathrm{l}$ is a hydrodynamic limiting value of the generalized
heat conductivity coefficient $\lambda^\mathrm{l}(\vec{k},\omega)$ defined via elements of the
matrix~(\ref{math/3.46}). $c_{p}$ and $c_{v}$ are the $\vec{k} \to 0$ limit values of the generalized heat capacity at constant
pressure $c_{p}(k)$ and at constant volume $c_{v}(k)$, respectively.
\emph{Two sound modes}
\begin{eqnarray}
\label{math/3.65}
z_{\pm}(k)=\pm\, i ck-\Gamma k^2, \qquad c=\frac{\gamma}{\beta m S(0)}, \qquad \Gamma=\frac{1}{2}\left[\left(\gamma-1\right)D_\mathrm{T}+\eta^\mathrm{L}\right],
\end{eqnarray}
where $c$ is the sound velocity in the system with the potential of interaction~(\ref{math/3.48}), $\gamma={c_p}/{c_v}$, and $S(0)$ is a limiting value of a static structure factor $S(k)$ of the system with potential~(\ref{math/3.48}).
$\Gamma$ is the sound damping coefficient.
\begin{eqnarray}
\label{math/3.66}
\eta^\mathrm{L}=\left(\frac{4}{3}\eta^{\perp}+\eta^\mathrm{b}\right)/mn, \qquad
\eta^\mathrm{b}=\eta^\mathrm{b}_\mathrm{E}+\eta^\mathrm{b}_\mathrm{l}, \qquad
\eta^{\perp}=\eta^{\perp}_\mathrm{E}+\eta^{\perp}_\mathrm{l}
\end{eqnarray}
$\eta^\mathrm{L}$ is the longitudinal viscosity defined via the bulk viscosity $\eta^\mathrm{b}$ and the shear viscosity $\eta^{\perp}$ coefficients, respectively. $\eta^{\perp}_\mathrm{E}$ is the shear viscosity in Enskog theory, and $\eta^{\perp}_\mathrm{l}$ is the value of the generalized shear viscosity coefficient $\eta^{\perp}_\mathrm{l}(\vec{k},\omega)$ [defined via elements of the matrix~(\ref{math/3.46})] calculated in the hydrodynamic limit
\begin{eqnarray}
\label{math/3.67}
\eta^{\perp}_\mathrm{l}=\lim_{\vec{k}\rightarrow 0,\omega\rightarrow 0}\eta^{\perp}_\mathrm{l}(\vec{k},\omega), \qquad
\eta^{\perp}_\mathrm{l}(\vec{k},\omega)=\frac{mn}{\beta}\frac{1}{k^{2}}
{\Sigma^{\prime\prime\perp}_{jj}}(\vec{k},\omega).
\end{eqnarray}
$\Sigma^{\perp}_{jj}(\vec{k},z)$ is the transverse component of the generalized transport kernel $\Sigma_{jj}(\vec{k},z)$, where the wave vector $\vec{k}$ is directed along the $0Z$ axis.
The longitudinal  viscosity coefficient $\eta^\mathrm{L}_\mathrm{l}$ is calculated in the
hydrodynamic limit
\begin{eqnarray}
\label{math/3.68}
\eta^\mathrm{L}_\mathrm{l}=\lim_{\vec{k}\rightarrow 0,\omega\rightarrow 0}\eta^\mathrm{L}_\mathrm{l}(\vec{k},\omega), \qquad
\eta^\mathrm{L}_\mathrm{l}(\vec{k},\omega)=\frac{mn}{\beta}\frac{1}{k^{2}}
{\Sigma^{\prime\prime\mathrm{L}}_{jj}}(\vec{k},\omega),
\end{eqnarray}
with $\eta^\mathrm{L}_\mathrm{l}(\vec{k},\omega)$ being the generalized longitudinal viscosity coefficient  defined via longitudinal components of the generalized transport kernel
$\Sigma_{jj}(\vec{k},z)$. \emph{Two transverse modes} with the eigenvalues given by
\begin{eqnarray}
\label{math/3.69}
z_{\nu}(k)=-\nu k^2, \qquad
\nu=\frac{\eta^{\perp}_\mathrm{E}+\eta^{\perp}_\mathrm{l}}{nm}.
\end{eqnarray}
$\nu$ is the kinematic viscosity for the system with the potential of interaction~(\ref{math/3.48}).

In the limit $k\sigma\gg1$, we obtain a
\emph{diffusion mode}, with the eigenvalue
\begin{eqnarray}
\label{math/3.70}
z_\mathrm{D}(k)=-D_\mathrm{E}k^2,
\end{eqnarray}
which is the same as in the Enskog theory.

As we can see from the above expressions, presence of the
long-range part in the potential of interaction entails a
renormalization of all the damping coefficients in the collective modes
spectrum. In particular, contributions related to long-range
potential appear in heat and sound modes as well as in transverse modes.
Nevertheless, diffusion mode remains unchanged.

When obtaining the spectrum of collective excitations based on the transport equations (\ref{math/3.49}), (\ref{math/2.12}) in the limit $|\vec{k}|\rightarrow 0$, $\omega \rightarrow 0$ we have not specificated the form
of the long-range potential of interaction $\Phi^\mathrm{l}(|\vec{r}_{ij}|)$.
Obviously, for every $|\vec{k}|$ and $\omega$ from the whole range of their values, the contributions from short-range and long-range parts of potential of interaction
into kinetic and hydrodynamic processes will be different and it is difficult to solve this problem in general. The kinetic processes connected with the particles scattering will be affected by both short- and long-range parts of interaction potential. Obviously, the long-wavelength hydrodynamic processes will be affected by the long-range part of potential $\Phi^\mathrm{l}(|\vec{r}_{ij}|)$ sufficiently, in particular, when the latter can be presented
via the Fourier-components of the particles number density (\ref{math/3.291}).
In order to evaluate the contribution of the long-range part into the collision integral
$\int d\vec{p}'\int_{-\infty}^t d t'e^{\varepsilon(t-t')}
\varphi_{nn}^\mathrm{l}(\vec{k};\vec{p},\vec{p}';t,t')f_{\vec{k}}(\vec{p}';t')$
we can employ an approach used to obtain the Enskog-Landau kinetic equation \cite{zub6} in the second order in long-range interaction. For our case of weakly nonequilibrium processes, we obtain:
\begin{eqnarray}
\label{math/3.701}\lefteqn{
\int d\vec{p}'\int_{-\infty}^t d t'e^{\varepsilon(t-t')}
\varphi_{nn}^\mathrm{l}(\vec{k};\vec{p},\vec{p}';t,t')f_{\vec{k}}(\vec{p}';t')=}
\\&&-n \vec{F}(\vec{k};t)\cdot\frac{\partial}{\partial
\vec{p}}f_{\vec{k}}(\vec{p};t)+n\int d\vec{v}'\int d\varphi \int_{0}^{\infty}db|\vec{q}|b\left[f_{0}(\vec{v}'^{*})f_{\vec{k}}(\vec{v};t)-f_{0}(\vec{v})f_{\vec{k}}(\vec{v}'^{*};t)\right],
\nonumber
\end{eqnarray}
where $\vec{F}(\vec{k};t)$ is the Fourier transform of the function $\vec{F}(r;t)=\frac{\partial}{\partial r}\Phi^\mathrm{l}(r)g_{2}(r;t)$, $b$ is the impact parameter, $\varphi$ is the azimuthal angle of scattering. $\vec{v}'^{*}$, $\vec{v}^{*}$ are the particle velocities  after collision:
\[
\vec{v}^{*}=\vec{v}+\Delta \vec{v},\qquad  \vec{v}'^{*}=\vec{v}'+\Delta \vec{v},
\qquad \Delta \vec{g}^{*}=\vec{g}^{*}-\vec{g}=2\Delta \vec{v}, \qquad
\vec{g}^{*}=\vec{v}'^{*}-\vec{v}^{*},
\]
\[
\Delta \vec{g}^{*}=\frac {1}{\mu^{*}|\vec{g}|}\int^{\infty_{-\infty}}d\xi \left(\frac{\partial}{\partial \vec{r}}\Phi^\mathrm{l}(r)\right)g_{2}(r;t)\Bigg|_{r=\sqrt{b^{2}-\xi^{2}}} \ ,
\]
$\xi$ is the distance along centreline of cylinder of two particles scattering, $\mu^{*}$ is the reduced mass and $g_{2}(r;t)$ is the two-particle quasiequilibrium coordinate distribution function. Herewith, it is assumed that scattering angles of particles a small, namely, $\chi^{*}(b, \vec{g})= {|\Delta \vec{g}^{*}|}/{|\vec{g}|}\ll 1$.
Thereby, it is possible to take into account an effect of the long-range part of the interaction potential in the region of particles scattering with a transfer of momentum.

Another important contribution of long-range part of interaction potential $\Phi^\mathrm{l}(|\vec{r}_{ij}|)$ is connected with transport kernels (\ref{math/3.24})--(\ref{math/3.25}), the potential part of the generalized heat conductivity coefficient $\bar{\varphi}_{\dot{h}\dot{h}}^{(0)}(\vec{k};t,t')$ as well as the coefficients
$\bar{\varphi}_{\jmath\dot{h}}^{(0)}(\vec{k};t,t')$ and $\bar{\varphi}_{\dot{h}\jmath}(\vec{k};t,t')$ describing cross-correlations between viscous and heat processes. In particular, to calculate $\bar{\varphi}_{\dot{h}\dot{h}}^{(0)}(\vec{k};t,t')$ it is necessary to reveal the action of Liouville operator onto $\hat{h}_{\vec{k}}^{int}$.
Taking into account (\ref{math/3.291}) with contribution from short-range interaction we obtain
\begin{eqnarray}
\label{math/3.2911}
\dot{\hat{h}}_{\vec{k}}^{int}=&-&i\vec{k}\cdot \sum_{i\neq j} \frac{\vec{p}_{i}}{m}\Phi^\mathrm{sh}(|\vec{r}_{ij}|) e^{-i\vec{k}\cdot\vec{r}_{i}}\\
&+&\frac{1}{2m}\sum_{\vec{q}}\nu (q)
\left[ -i(\vec{q}+\vec{k})\cdot\hat{\vec{j}}_{\vec{q}+\vec{k}}
\hat{n}_{-\vec{q}}+i\vec{q}\cdot\hat{n}_{\vec{q}+\vec{k}}\hat{\vec{j}}_{-\vec{q}}\right]\nonumber\\
&-&\sum_{\vec{q}}\nu (q)S_{3}(\vec{q}+\vec{k},-\vec{q},-\vec{k})S_{2}^{-1}(\vec{q})i\vec{k}\cdot\hat{\vec{j}}_{-\vec{k}}\,.
\end{eqnarray}
Taking into account the structure of $\dot{\hat{h}}_{\vec{k}}^{int}$  for $\bar{\varphi}_{\dot{h}\dot{h}}^{(0)}(\vec{k};t,t')$ we obtain:
\begin{equation} \label{math/3.2912}
\bar{\varphi}_{\dot{h}\dot{h}}^{(0)}(\vec{k};t,t')=\bar{\varphi}_{\dot{h}\dot{h}}^{sh,sh}(\vec{k};t,t')+\bar{\varphi}_{\dot{h}\dot{h}}^{sh,l}(\vec{k};t,t')+
\bar{\varphi}_{njj}^{ll}(\vec{k};t,t')+\bar{\varphi}_{njnj}^{ll}(\vec{k};t,t'),
\end{equation}
where, in particular,
\[
\bar{\varphi}_{\dot{h}\dot{h}}^{sh,sh}(\vec{k};t,t')
=k^{2}\langle \sum_{ij} \frac{\vec{p}_{i}}{m}\Phi^\mathrm{sh}(|\vec{r}_{ij}|) e^{-\vec{k}\vec{r}_{i}}T_0(t,t')
\sum_{lj} \frac{\vec{p}_{l}}{m}\Phi^\mathrm{sh}(|\vec{r}_{lj}|) e^{\vec{k}\vec{r}_{l}}\rangle_0
\Phi_{hh}^{-1}(\vec{k})
\]
\begin{eqnarray} \label{math/3.293}
&&\bar{\varphi}_{njnj}^{ll}(\vec{k};t,t')=\left(\frac{1}{2m}\right)^{2}\sum_{\vec{q}\vec{q}'}\nu (q)\nu (q')\\
&&\times\Big[-(\vec{q}+\vec{k})\cdot\langle \hat{\vec{j}}_{\vec{q}+\vec{k}}\hat{n}_{-\vec{q}}T_0(t,t')\hat{\vec{j}}_{\vec{q}'-\vec{k}}
\hat{n}_{-\vec{q}'}\rangle_{0}\cdot(\vec{q}'-\vec{k})
+(\vec{q}+\vec{k})\cdot\langle \hat{\vec{j}}_{\vec{q}+\vec{k}}\hat{n}_{-\vec{q}}T_0(t,t')\hat{n}_{-\vec{q}'-\vec{k}}\hat{\vec{j}}_{-\vec{q}'}\rangle_{0}\cdot \vec{q}'\nonumber\\
&&+\vec{q}\cdot \langle \hat{n}_{\vec{q}+\vec{k}}\hat{\vec{j}}_{-\vec{q}}T_0(t,t')
\hat{\vec{j}}_{\vec{q}'-\vec{k}}\hat{n}_{-\vec{q}'}\rangle_{0}\cdot(\vec{q}'-\vec{k})
-\vec{q}\cdot \langle \hat{n}_{\vec{q}+\vec{k}}\hat{\vec{j}}_{-\vec{q}}T_0(t,t')
\hat{n}_{-\vec{q}'-\vec{k}}\hat{\vec{j}}_{-\vec{q}'}\rangle_{0}\cdot \vec{q}'\Big]
\Phi_{hh}^{-1}(\vec{k})\nonumber
\end{eqnarray}%
Here, the time correlation function
$\bar{\varphi}_{\dot{h}\dot{h}}^{sh,sh}(\vec{k};t,t')$ contributes from the short-range interaction and depends on the model for potential $\Phi^\mathrm{sh}(|\vec{r}_{ij}|)$. $\bar{\varphi}_{\dot{h}\dot{h}}^{sh,l}(\vec{k};t,t')$ is the time correlation function describing cross-correlations between contributions from the short- and long-range parts of the potential of interaction. $\bar{\varphi}_{njj}^{ll}(\vec{k};t,t')$ and $\bar{\varphi}_{njnj}^{ll}(\vec{k};t,t')$ are the time correlation functions of the third and the fourth order in dynamic variables $\hat{\vec{j}}$, $\hat{n}\hat{\vec{j}}$  with the evolution operator $T_0(t,t')$.
Investigating the time correlation functions for $\bar{\varphi}_{njnj}^{ll}(\vec{k};t,t')$ an approximation of the mode coupling type \cite{Get} via the time correlation functions  $\Phi_{nn}\Phi_{jj}+\Phi_{nj}\Phi_{jn}$ can be used.
This issue needs a more detailed investigation within a certain model for the potential of interaction between particles.

\section{Conclusions}

In summary, in the present paper we consider the set of equations for the nonequilibrium one-particle distribution function and the potential part of the enthalpy density
obtained within the framework of a consistent description of kinetics and hydrodynamics in dense gases and liquids. In this set of equations the collision integral of the kinetic equation has the Fokker-Planck form with the generalized friction coefficient in momentum
space. It was shown that in this approach the viscosity processes are described in terms of the generalized diffusion and friction coefficients in the phase space. Applying the procedure of projecting onto the moments of the nonequilibrium distribution function to the
equations~(\ref{math/2.12}),~(\ref{math/3.49}) we obtain a set of
equations for extended set of variables $\{n_{\vec{k}}(z), \ \vec{\jmath}_{\vec{k}}(z), \  h_{\vec{k}}^{kin}(z), \ h_{\vec{k}}^{int}(z)\}$ and perform the transition to the transport equations of molecular hydrodynamics.

Within the framework of consistent description of kinetic and hydrodynamic processes we considered a set of kinetic equations for the potential of interaction of the
system presented by the sum of hard spheres potential
$\Phi^\mathrm{hs}(|\vec{r}_{ij}|)$ and a certain smooth one
$\Phi^\mathrm{l}(|\vec{r}_{ij}|)$. In this case, we separated the
Enskog-Boltzmann  collision integral describing a collision
dynamics at short distances from the collision integral of the
kinetic equation for the nonequilibrium distribution function.
Applying the procedure of projecting and passing to equations of molecular hydrodynamics we obtain a set of equations which allow us to investigate a spectrum of collective excitations in the limits $k\sigma\ll1$ and $k\sigma\gg1$. We showed that, besides the
contribution from the hard spheres potential, all hydrodynamic modes contain contributions from the long-range part of potential. These contributions make the
damping coefficients closer to the ones known from the hydrodynamic
theory. Here, we formally presented the contribution from the
long-ranged part of potential, since the latter, for example the
Coulomb one, will contribute into the transport kernels~(\ref{math/2.15}).
Moreover, we can separate the linearized
Landau-like collision integral (\ref{math/3.701}) describing pair collisions in
$\varphi_{nn}(\vec{k};\vec{p},\vec{p}';t,t')$, while
$\varphi_{hn}(\vec{k};\vec{p};t,t')$,
$\varphi_{nh}(\vec{k};\vec{p};t,t')$,
$\varphi_{hh}(\vec{k};\vec{p};t,t')$ take into account collective
Coulombic interactions. Evidently, calculation of the elements
(\ref{math/3.47}) of matrix $\tilde{\Pi}(\vec{k};z)$ will depend
on the model of time dependence (exponential, Gaussian etc.) for
transport kernels (\ref{math/2.15}). When a spectrum of
collective excitations is known, a whole set of time correlation
functions can be investigated. In particular, it makes possible to
investigate the behaviour of the dynamic structure factor and, in the
case of potential~(\ref{math/3.48}), to separate a contributions
from the hard spheres potential and the long-range part of
potential in it.

\appendix

\renewcommand{\theequation}{A.\arabic{equation}}

\section*{Appendix}

Here we present some correlation functions appearing in section~\ref{sec:3}.
Correlation functions $\bar{\varphi}_{nn}^{(1)}(\vec{k};\vec{p},\vec{p}';t,t')$ entering the equations (\ref{math/3.9}) has the following form:
\begin{eqnarray}
\label{math/3.16}
\lefteqn{\bar{\varphi}_{nn}^{(1)}(\vec{k};\vec{p},\vec{p}';t,t')=
\frac{\beta}{m}\vec{p}\cdot\vec{\Phi}_{Fh}(\vec{k})f_0(p)
\varphi_{h\jmath}(\vec{k};\vec{p}';t,t')\cdot\frac{i\vec{k}}{m}}
\\&&\mbox{}+-\frac{\vec{k}}{m}\cdot\vec{p}\varphi_{n\jmath}(\vec{k};\vec{p},\vec{p}';t,t')
\cdot\frac{\vec{k}}{m}
\frac{\beta}{m}\vec{p}\cdot\vec{\Phi}_{Fh}(\vec{k})f_0(p)
\varphi_{hF}(\vec{k};\vec{p}';t,t')\cdot\frac{\partial}{\partial\vec{p}'}\nonumber
\\&&\mbox{}+
\frac{i\vec{k}}{m}\cdot\vec{p}\varphi_{nF}(\vec{k};\vec{p},\vec{p}';t,t')
\cdot\frac{\partial}{\partial\vec{p}'}-\frac{\vec{k}}{m}\cdot\varphi_{\jmath
n}(\vec{k};\vec{p},\vec{p}';t,t') \frac{\vec{k}}{m}\cdot\vec{p}'-
\frac{\partial}{\partial\vec{p}}\cdot\vec{p}\varphi_{Fn}(\vec{k};\vec{p},\vec{p}';t,t')
\frac{i\vec{k}}{m}\cdot\vec{p}' \nonumber
\\&&\mbox{}-\frac{\beta}{m}\vec{p}\cdot\vec{\Phi}_{Fh}(\vec{k})f_0(p)
\varphi_{hn}(\vec{k};\vec{p}';t,t')\frac{i\vec{k}}{m}\cdot\vec{p}'+ \frac{\vec{k}}{m}\cdot\vec{p}
\varphi_{nn}^{(0)}(\vec{k};\vec{p}';t,t')\frac{\vec{k}}{m}\cdot\vec{p}'.\nonumber
\end{eqnarray}
It contains the following correlation functions
\begin{align}
\label{math/3.17}
&\varphi_{nn}^{(0)}(\vec{k};\vec{p},\vec{p}';t,t')=
\langle\hat{n}_{\vec{k}}(\vec{p})T_0(t,t')
\hat{n}_{-\vec{k}}(\vec{p}')\rangle_0,
& &\varphi_{hn}(\vec{k};\vec{p}';t,t')=
\langle\hat{h}^{int}_{\vec{k}}T_0(t,t')
\hat{n}_{-\vec{k}}(\vec{p}')\rangle_0,\nonumber\\
&\varphi_{h\jmath}(\vec{k};\vec{p}';t,t')=
\langle\hat{h}^{int}_{\vec{k}}T_0(t,t')
\hat{\vec{\jmath}}_{-\vec{k}}(\vec{p}')\rangle_0,
& &\varphi_{hF}(\vec{k};\vec{p}';t,t')=
\langle\hat{h}^{int}_{\vec{k}}T_0(t,t')
\hat{\vec{F}}_{-\vec{k}}(\vec{p}')\rangle_0,\\
&\varphi_{n\jmath}(\vec{k};\vec{p},\vec{p}';t,t')=
\langle\hat{n}_{\vec{k}}(\vec{p})T_0(t,t')
\hat{\vec{\jmath}}_{-\vec{k}}(\vec{p}')\rangle_0,
& &\varphi_{nF}(\vec{k};\vec{p},\vec{p}';t,t')=
\langle\hat{n}_{\vec{k}}(\vec{p})T_0(t,t')
\hat{\vec{F}}_{-\vec{k}}(\vec{p}')\rangle_0,\nonumber\\
&\varphi_{Fn}(\vec{k};\vec{p},\vec{p}';t,t')=
\langle\hat{\vec{F}}_{\vec{k}}(\vec{p})T_0(t,t')
\hat{n}_{-\vec{k}}(\vec{p}')\rangle_0,
& &\varphi_{hn}(\vec{k};\vec{p}';t,t')=
\langle\hat{h}^{int}_{\vec{k}}T_0(t,t')
\hat{n}_{-\vec{k}}(\vec{p}')\rangle_0,\nonumber
\end{align}
constructed on the basic set of dynamic variables
$\hat{n}_{\vec{k}}(\vec{p})$, $\hat{h}^{int}_{\vec{k}}$ along with
the Fourier-components of the momentum density
$\hat{\vec{\jmath}}_{\vec{k}}(\vec{p})$ and the force
$\vec{F}_{\vec{k}}(\vec{p})$ in momentum space. Moreover,
$\hat{n}_{\vec{k}}(\vec{p})$,
$\hat{\vec{\jmath}}_{\vec{k}}(\vec{p})$ and
$\vec{F}_{\vec{k}}(\vec{p})$ are connected by the equation of
motion (\ref{math/3.1}).


The  correlation functions (\ref{math/3.17}) enter
into $\varphi_{n\jmath}^{(2)}(\vec{k};\vec{p};t,t')$ from equation (\ref{math/3.8}):
\begin{eqnarray}
\label{math/3.18}\lefteqn{\varphi_{n\jmath}^{(2)}(\vec{k};\vec{p};t,t')
=\frac{i\vec{k}}{m}\cdot\vec{p}\varphi_{nh}(\vec{k};\vec{p};t,t')
\frac{\beta}{mn}\vec{\Phi}_{Fh}(\vec{k})+
\frac{\beta}{m}\vec{p}\cdot\vec{\Phi}_{Fh}(\vec{k})f_0(p)
\varphi_{hh}(\vec{k};t,t')\frac{\beta}{mn}\vec{\Phi}_{Fh}(\vec{k})}\nonumber
\\&&\mbox{}-\frac{i\vec{k}}{m}\cdot\varphi_{\jmath h}(\vec{k};\vec{p};t,t')
\frac{\beta}{mn}\vec{\Phi}_{Fh}(\vec{k}) +\frac{\partial}{\partial
\vec{p}}\cdot\varphi_{Fh}(\vec{k};\vec{p};t,t')
\frac{\beta}{mn}\vec{\Phi}_{Fh}(\vec{k}),
\qquad\qquad\qquad\qquad\qquad
\end{eqnarray}
where
\begin{eqnarray}
\label{math/3.19}
&\varphi_{\jmath h}(\vec{k};\vec{p};t,t')
=\langle\hat{\vec{\jmath}}_{\vec{k}}(\vec{p})T_0(t,t')
\hat{h}_{-\vec{k}}^{int}\rangle_0,
&\varphi_{Fh}(\vec{k};\vec{p};t,t')
=\langle\vec{F}_{\vec{k}}(\vec{p})T_0(t,t')
\hat{h}_{-\vec{k}}^{int}\rangle_0,  \\
&\varphi_{hh}(\vec{k};t,t')=\langle\hat{h}^{int}_{\vec{k}}
T_0(t,t')\hat{h}^{int}_{-\vec{k}}\rangle_0\nonumber
\end{eqnarray}
the correlation functions of the Fourier-components of
densities of the potential part of enthalpy, momentum
$\hat{\vec{\jmath}}_{\vec{k}}(\vec{p})$ and force
$\vec{F}_{\vec{k}}(\vec{p})$ in impulse space.


%
The correlation functions entering the equation (\ref{math/3.21}) have the following structure
\begin{align}
\label{math/3.22}
&\bar{\varphi}_{\jmath\dot{h}}(\vec{k};\vec{p};t,t')
=\langle\hat{\vec{\jmath}}_{\vec{k}}(\vec{p})T_0(t,t')
\dot{\hat{h}}_{-\vec{k}}^{int}\rangle_0\Phi_{hh}^{-1}(\vec{k}),\nonumber\\
&\bar{\varphi}_{F\dot{h}}(\vec{k};\vec{p};t,t')
=\langle{\vec{F}}_{\vec{k}}(\vec{p})T_0(t,t')
\dot{\hat{h}}_{-\vec{k}}^{int}\rangle_0\Phi_{hh}^{-1}(\vec{k}),\nonumber\\
&\bar{\varphi}_{h\dot{h}}(\vec{k};t,t')
=\langle\hat{h}^{int}_{\vec{k}}(\vec{p})T_0(t,t')
\dot{\hat{h}}_{-\vec{k}}^{int}\rangle_0\Phi_{hh}^{-1}(\vec{k}),\nonumber\\
&\bar{\varphi}_{n\dot{h}}(\vec{k};t,t')
=-\langle\dot{\hat{n}}_{\vec{k}}(\vec{p})T_0(t,t')
{\hat{h}}_{-\vec{k}}^{int}\rangle_0\Phi_{hh}^{-1}(\vec{k})
=\frac{i\vec{k}}{m}\cdot\bar{\varphi}_{\jmath
h}(\vec{k};\vec{p};t,t')-\frac{\partial}{\partial
\vec{p}}\cdot\bar{\varphi}_{Fh}(\vec{k};\vec{p};t,t'),\nonumber\\
%
&\bar{\varphi}_{\jmath h}(\vec{k};\vec{p};t,t')
={\varphi}_{\jmath h}(\vec{k};\vec{p};t,t')\Phi_{hh}^{-1}(\vec{k}),\nonumber\\
&\bar{\varphi}_{Fh}(\vec{k};\vec{p};t,t')
={\varphi}_{Fh}(\vec{k};\vec{p};t,t')\Phi_{hh}^{-1}(\vec{k})%
\end{align}
and they are the normalized time correlation functions.

\end{document}